# Resolving the Electron Plume within a Scanning Electron Microscope

Francis M. Alcorn*[1], Christopher Perez[1,2], Eric J. Smoll[1], Lauren Hoang[3], Frederick Nitta[3,4], Andrew J. Mannix[4], A. Alec Talin[1], Craig Y. Nakakura[5], David W. Chandler*[1], Suhas Kumar*[1]

[1] *Sandia National Laboratories, Livermore, CA, USA*
[2] *Department of Mechanical Engineering, Stanford University, Stanford, CA, USA*
[3] *Department of Electrical Engineering, Stanford University, Stanford, CA, USA*
[4] *Department of Materials Science and Engineering, Stanford University, Stanford, CA, USA*
[5] *Sandia National Laboratories, Albuquerque, NM, USA*

**Abstract**

Scanning electron microscopy (SEM), a century-old technique, is today a ubiquitous method of imaging the surface of nanostructures. However, most SEM detectors simply count the number of secondary electrons from a material of interest, and thereby overlook the rich material information contained within them. Here, by simple modifications to a standard SEM tool, we resolve the momentum and energy information of secondary electrons by directly imaging the electron plume generated by the electron beam of the SEM. Leveraging these spectroscopic imaging capabilities, our technique is able to image lateral electric fields across a prototypical silicon p-n junctions and to distinguish differently doped regions, even when buried beyond depths typically accessible by SEM. Intriguingly, the sub-surface sensitivity of this technique reveals unexpectedly strong surface band bending within nominally passivated semiconductor structures, providing useful insights for complex layered component designs, in which interfacial dynamics dictate device operation. These capabilities for non-invasive, multi-modal probing of complicated electronic components are crucial in today's electronic manufacturing but is largely inaccessible even with sophisticated techniques. These results show that seemingly simple SEM can be extended to probe complex and useful material properties.

**Significance Statement:**

Scanning electron microscopy (SEM) is a ubiquitous tool in materials characterization, particularly for imaging semiconductor devices. By a simple modification of a conventional SEM, we can resolve secondary electrons by their energies and momenta, and uncover information about a semiconductor p-n junction that is otherwise hidden in typical SEM. Most notably, we have demonstrated that momentum-resolved secondary electrons can be used to directly image electric fields across a device. These findings highlight the capability to expand the utility of SEM, an already widespread and established technique, for non-invasive characterization of semiconductor devices.

**Introduction**

Modern computer technology relies upon nanoscale device patterning, which necessarily requires techniques for characterization of structures at nanometer length scales. Scanning electron microscopy (SEM), with nanometer spatial resolution and requiring minimally invasive sample preparation for imaging native-state semiconductor devices, is a ubiquitous tool for such measurements. In SEM, images are produced by scanning a focused probe of electrons with energies ~1-30 keV across a surface, generating both scattered and secondary electrons that are collected, counted, and plotted versus electron beam position to produce an image. Typically, these generated electrons are measured with an Everhart Thornley detector (ETD),[1] which uses a positive bias to pull electrons onto a scintillator and quantifies the generated signal using a photomultiplier tube (PMT). While fully adequate for routine SEM imaging, ETDs have a drawback in that they cannot collect information stored in the trajectories of the generated electrons, such as their energies and emission angles, which can provide useful information on the electronic structure of a sample.[2] As such, SEM can provide a very primitive characterization of semiconductor devices, while probing of non-trivial information (such as momentum-resolved electronic structure) or structures (such as deep buried interfaces) requires either destructive or bespoke instruments such as angle-resolved photoelectron spectroscopy (ARPES)[5] or focused ion beam milling for transmission electron microscopy analysis (FIB-TEM).[3,4] In today's world of semiconductor devices employing heterogeneous and multi-layer integration, and requiring complicated processing lines, the lack of in-line and non-destructive probing leads to costly and time-consuming failure analysis.[6-8]

Several types of electrons are generated by an SEM electron beam; secondary electrons (SEs), backscattered electrons (BSEs), and Auger electrons.[2] SEs are the most prevalent and have a distribution of energies from 0 to over 50 eV with a peak value around ~5-10 eV, and a decreasing tail at higher energies. The exact SE energy distribution of SEs varies with experimental parameters including electron beam energy, but also depends on sample surface properties, particularly surface potentials.[2,9,10] Consequently, SE energy spectra have been exploited for measuring work functions and electronic densities of surface states.[10-12] Additionally, energy and angular distributions of SEs can inform on crystal structure of samples,[13,14] or their specific trajectories can be used to image domains on a magnetic sample.[15,16] Auger electrons are another class of electrons generated by the electron beam irradiation. They have energies on the order of ~100 eV that are dictated by the core electron orbitals of the constituent elements, and are useful for elemental identification and mapping.[2,17] Lastly, the primary electron beam can be elastically backscattered in a process known as electron backscattered diffraction (EBSD) for sample crystallography.[18,19] Without any way to resolve electron energies or angular distributions, all this information pertaining to the sample structure is lost when using an ETD.

To address this shortcoming, we designed a detector that enables energy and momentum resolution of SEs within an SEM tool. Briefly, an electrostatic bias is used to project the SE distribution generated from the surface by the electron beam—herein referred to as the 'electron plume'—onto a two-dimensional (2D) screen with the two planar axes related to electron energy and momenta. As a model system, we investigate simple lateral p-n junctions, revealing how lateral electric fields can deflect the electron plume resulting in edge contrast at the p-n interfaces That is not detected by a typical ETD nor other complementary techniques, and the signal is observable even through a thick passivating oxide layer. Additionally, initial results suggest that this detector can be applied for chemical discrimination of materials by recording the energy distribution of the secondary electrons. These results showcase the potential of this detector for non-invasive, quantitative characterization of devices even under a surface passivating layer, as well as a tool for studying both below- and above- surface interfacial physics in such samples.

**Detector premise and design**

SEM in principle operates by scanning a focused electron probe across a surface and recording an electron signal as a function of position of the electron probe during its raster. Conventionally, an ETD is used to simply count electrons at each position (Fig. 1a), thereby missing out on the dynamical properties (energy and angle distributions) of the generated SEs, which we herein refer to as the 'SE plume'. To investigate these dynamical properties, we designed a detector that (Fig. 1b) uses an electrostatic potential to project the three dimensional (3D) SE plume onto a 2D position sensitive detector, referred to herein as the viewing screen, where in one dimension, the $z$-axis, electrons are separated by energy with higher energy electrons impacting higher on the screen, and in the other dimension, the $x$-axis, electrons are separated by angle or lateral momentum, impacting the screen at farther distances from the main axis of the plume with increasing emission angle.

A full schematic of the detector, including all electron optics and the components of the viewing screen, is provided in Fig. 1c (photographs in Fig. S1). Briefly, the SE plume is pushed up away from the sample by applying a negative bias of ~100 – 200 V to the sample electrode, on which the sample wafer is placed. The sample wafer and electrode are enclosed within a grounded, rectangular 'shroud' electrode with a precision air slit (Edmund Optics) aperture (0.2 or 0.5 mm by 3 mm rectangle) providing an opening in the shroud through which the electron beam can access the sample wafer, and which allows generated SEs to escape for detection. The ~100 – 200 V change in potential from the sample to shroud electrode accelerates the SEs so that their energies upon exiting the slit are the same order of magnitude, thereby mitigating distortions in the SE plume arising from electron optics acting on SEs with widely varying energies. Upon exiting the slit on the shroud, a repeller electrode, biased at negative 3000 – 5000 V, pushes the plume into a grounded mu metal time-of-flight (ToF) tube with Cu mesh on the front opening, projecting the electrons onto a position sensitive viewing screen consisting of a pair of microchannel plates (MCPs), each biased to ~700 – 900 V, and a phosphor screen, biased to ~3000 – 4000 V. In addition to the sample, shroud, and repeller electrodes, side electrodes control the lateral position of the SE plume by biasing the front electrode relative to the grounded back electrode for centering the plume on the viewing screen. These side electrodes further help to mitigate the effects of stray electromagnetic fields. After the electrons are amplified by the MCPs and converted into photons by the phosphor, signal is collected and digitized using an appropriate photodetector. A digital camera was used to collect images of the SE plume on the viewing screen and a photomultiplier tube (PMT) was used as a fast photodetector synchronized to the electron beam raster for producing SEM micrographs from the SE plume. In all, the detector, effectively composed of machined stainless-steel electrodes, a mu metal ToF tube, an electron-responsive light emitter, and a simple digital camera or other routine photodetector, offers an inexpensive avenue towards unlocking the information stored within a SE distribution in SEM. Further experimental details, including electron optics geometries (Table S1) are provided in supporting information.

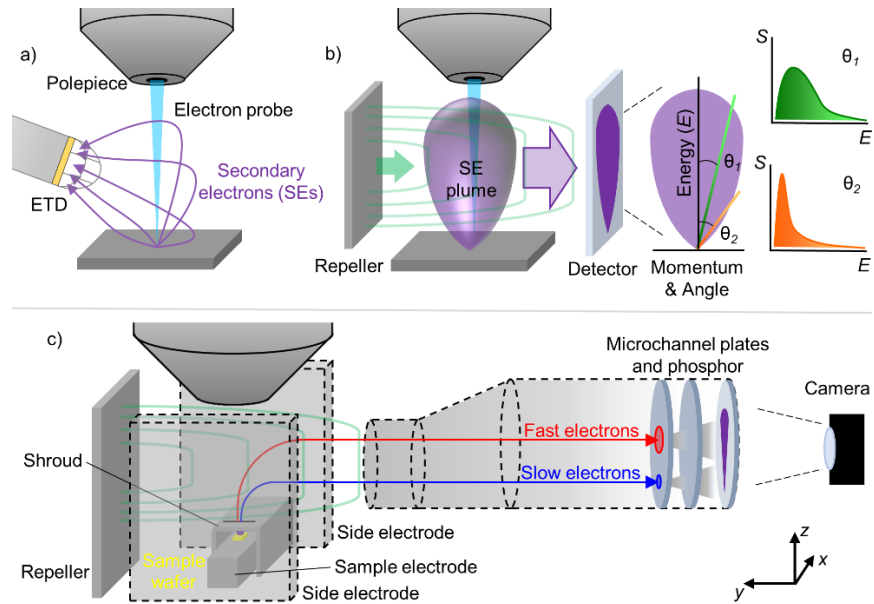

**Figure 1| Premise and schematic of detector.** SEM utilizes a focused electron probe (primary electron beam) to generate SEs, and other electrons, that are collected and quantified to produce an image. **a** These SEs contain information related to sample structure which is lost in conventional SEM imaging with an ETD, which cannot discriminate electron trajectories. **b** The principle of our energy- and momentum- resolved detector is to project the SE distribution (SE plume) onto a 2D viewing screen with axes related to electron energies ($z$) and momenta ($x$) using an electrostatic repeller electrode. By recording the SE plume in 2D, rather than using a zero-dimensional detector to simply count electrons such as an ETD, we can extract energy ($E$) and angular information from the SE plume that relates back to the electronic structure of the sample. Orange and green spectra, (signal or electron count, $S$, versus $E$) highlight the multidimensionality of the design for investigating SE distributions in angle and energies. **c** Schematic of the full detector; electron optics (sample electrode, shroud, side electrodes, and repeller) align and project the plume into a mu metal tube and onto the viewing screen composed of two MCPs and a phosphor screen. This amplifies the signal and converts it to light to be collected by a photodetector (camera or PMT). Electrons are spread over the detector due to their angular distribution, as denoted by the blue and red discs where electrons impact the MCPs in this schematic. Faster electrons experiencing larger spreads. Green lines in **b** and **c** schematically illustrate the equipotential lines of the detector in operation, with the green arrow in **b** denoting the direction electrons are repelled by the applied voltage. Arrows in the bottom right in **c** provide a real-space coordinate system for our detector.

This detector design was validated by SIMION, an ion-optics simulation program. Specifically, simulations of electron trajectories through our detector demonstrate the capability of separating electrons by energies in the *z*-dimension (Fig. S2) or control the impact heights electrons on the viewing screen using the sample (Fig. S3, S4), offering a way to optimize detector response for higher energy electrons (e.g., Auger electrons). Additionally, simulations suggest that slit geometry impacts the plume shape, which was also observed experimentally (Fig. S5), and are in use to understand the response of the detector using expected SE distributions (Fig. S12-S14).

**Imaging interfacial electric fields with momentum-resolved electrons**

To demonstrate the capabilities of this detector, we chose to investigate planar p-n junctions, the basic building block of many semiconductor devices. Differences in doping and surface band bending of p- and n-doped regions are known to result in changes in SE contrast (Fig. 2b).[20-25] Further, at the p-n junction, differences in work functions result in interfacial band bending and formation of a lateral built-in electric field (Fig. 2c), a dynamic that governs the operation of devices such as diodes. Thus, there are several phenomena of interest in these p-n junction samples that can be investigated using our detector: resolving effects of lateral electric fields at the p-n interfaces using SE momentum and comparison of energy distributions of SE plumes for p- and n-doped regions.

All experiments reported herein were performed with a primary electron beam energy of 10 keV. The p-n junction samples, illustrated in Fig. 2a, were made by patterning an Si wafer with p-doped (B, ~$10^{19}$ cm$^{-3}$) squares surrounded by n-doped (As, ~$10^{17}$ cm$^{-3}$) material and have a thermal oxide covering of ~3.5 nm thickness. An additional 50 nm SiO$_2$ was sputter coated over the sample to facilitate comparison between imaging with a thin and a thick oxide layer. Further, the chosen structure is largely flat to remove effects from topography on SE distributions. Details for synthesis can be found in a previous report.[26] Other experimental details are included in the supporting information.

Micrographs of p-doped Si squares surrounded by n-doped Si (Fig. 2d) were recorded by collecting signal from the central portion of the SE plume—thereby only collecting electrons with nearly vertical initial trajectories—revealing several different contrast mechanisms in our detector. First, there is a clear difference in relative electron counts for p- and n-doped Si, with higher counts (dark blue) for the p-doped material than the n-doped material (light green), reflecting the expected doping contrast that arises from the differences in band bending at the surface of p-doped and n-doped regions.[20-25] Specifically, the surface band bending in p(n)-doped regions decrease (increase) the energy barrier for SE emission, resulting in increased (decreased) SE yield (Fig. 2b). This contrast is also visible in ETD images of this feature, in which the p-doped square is brighter (higher electron counts) than the surrounding material (Fig. 2e). There is also a second contrast mechanism present in the micrograph recorded using the SE plume that is missing in the ETD images. Specifically, the p-doped square is surrounded by a low-electron-count (yellow) halo immediately at the p-n interface (Fig. 2d, red dashed box). This edge contrast is due to the lateral built-in electric field that governs interfacial dynamics in a p-n junction (Fig. 2f-h).

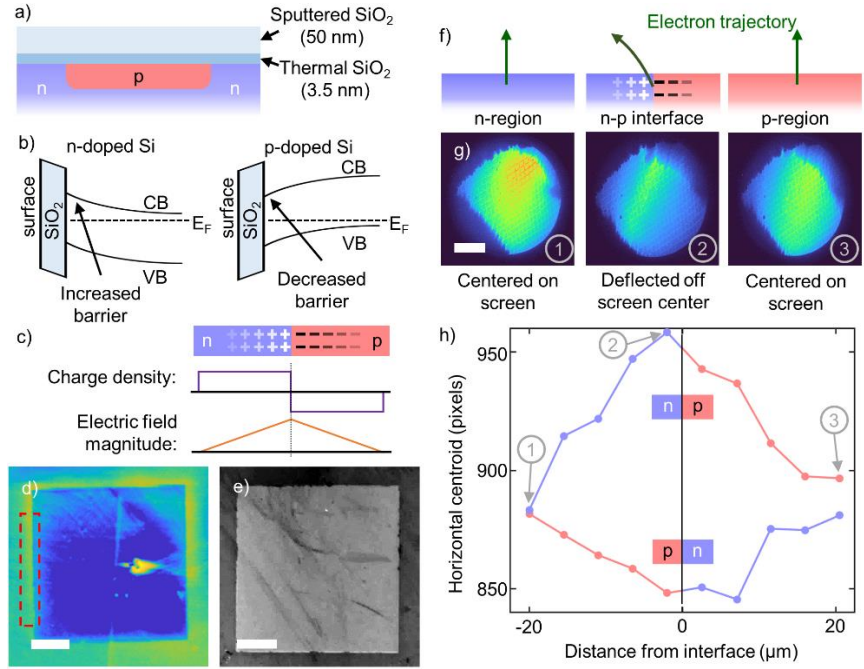

**Figure 2| Electric field contrast with momentum resolved SEs. a** Schematic of the sample used in these studies. The sample is composed of regions of p-doped Si surrounded by n-doped Si. A ~3.5 nm thermal oxide layer is on the surface. For studies of dynamics under a thick oxide layer discussed later, an additional 50 nm of $SiO_2$ was sputtered on top of the thermal oxide. **b** Band bending at the sample surface differs for p- and n-doped regions, which has effects on SE emission, resulting in the known dopant contrast in SEM of semiconductor materials. Specifically, SE emission is higher from p-doped material than n-doped material. **c** At a p-n junction, the built-in potential difference between work functions of the p- and n- doped materials results in charge carrier diffusion across the interface forming a space-charge region (known as the depletion region) and resulting in the built-in electric field. **d** The expected dopant contrast was seen in a micrograph of a p-doped square surrounded by n-doped material recorded using our electron detector, in which the square is darker blue (higher electron counts) than the surrounding wafer (green/fewer electron counts). An additional contrast mechanism is seen at the edges of the square (red dashed box), in which electron counts are reduced (yellow) immediately at the p-n interfaces. **e** Only dopant contrast is observed in a micrograph recorded using an ETD. **f** The edge contrast seen in **d** arises from interfacial electric fields at the p-n junctions which deflects the SE plume on the detector. This is seen in photographs of the SE plume **g**, with the SE plume deflecting to the left in the photograph at a p-n junction with the orientation shown in **f**. **h** The centroid position of the plume on the camera shifts near the p-n interface with opposite deflection directions for opposite orientations and returns roughly to the center of the screen far from the interface on either side. All images in this data set are provided in Fig. S7. Error bars denoting the standard deviation from peak fitting are present in this plot but are too small to see at this scale. Scale bars 25 μm in **d** and **e**, 2 cm or ~200 pixels in **g**. Photos in **g** correspond to schematics in **f** and numbered points in **h**.

These lateral electric fields physically deflect the SE plume and move it on the viewing screen (Fig. 2f-h). Due to formation of a depletion region at a p-n interface, positive charges in the n-doped region and negative charges in the p-doped region at the interface result in the SEs being deflected away from the p-doped material. This deflection was directly observed in photographs of the SE plume at various distances from a p-n interface (Fig. 2g,h). In regions far from an interface, there are minimal lateral electric fields just above the sample surface to interact with the SE plume, and thus the SE plume is roughly centered on the viewing screen when properly aligned (Fig. 2g, ① and ③). However, at the p-n interfaces there are substantial lateral electric fields, and the plume appears off-centered on the viewing screen (Fig. 2g, ②). This deflection was tracked as a function of position on a line profile (Fig. S7) across two p-n interfaces, one with the n-doped region on the left and the p-doped region on the right and a second with the opposite geometry, i.e. p- on the left and n- on the right (Fig. 2h). Centroid positions of the SE plume on the camera in the horizontal dimension were calculated by integrating the electron plume along the vertical dimension in the photograph and fitting the integrated vector to a mixed Gaussian/Lorentzian peak to capture lateral shifts in the SE plume due to the lateral electric fields. For both plots, the SE plume was roughly centered on the viewing screen, at a position on the camera of ~880 – 900 pixels, far (~20 μm) from the p-n interface, and deflected off-center near the interface, with the opposite interface geometries showing opposite directions of deflection on account of their oppositely oriented electric fields (Fig. 2h). The degree of the deflection experienced by electrons was different for different energies, with lower energy electrons being more significantly deflected by these lateral electric fields (Fig. S8), which also results in an apparent rotation of the SE plume on the viewing screen. In both cases, the electrons are deflected away from the p-doped regions, consistent with the effects of interfacial band bending from the built-in potential difference that forms a depletion region on either side of the interface (Fig. 2f). Altogether, these results point to a unique capability of our electron detector arising from its momentum resolution.

Specifically, we have demonstrated that momentum-resolved SEs can be used for directly studying lateral electric fields, such as those present at interfaces in devices. With proper calibration, the degree of SE plume deflection can be used for quantitative analysis of lateral electric fields, useful for understanding technologies utilizing interfacial electric fields in their functions, such as transistors, photovoltaics, or even electrocatalytic systems. Furthermore, the directionality of the deflection can be directly exploited for mapping electric or magnetic fields in a sample.

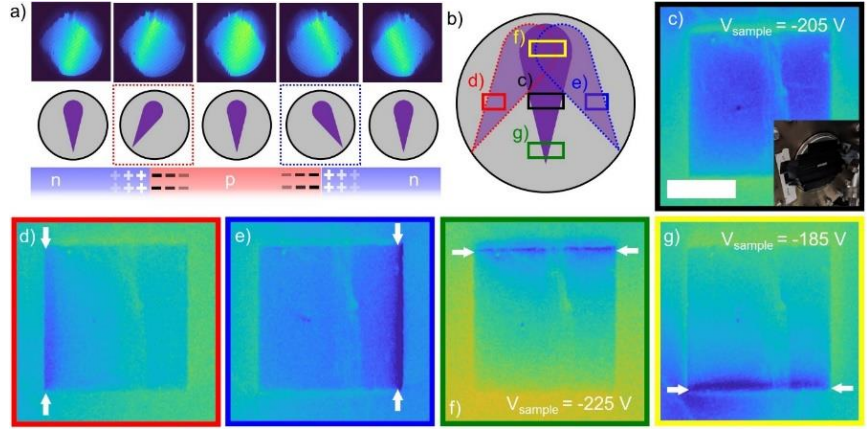

**Figure 3| Imaging p-n junction electric fields**. The deflection of the SE plume due to interfacial electric fields at p-n junctions **a** can be exploited for selectively imaging features by the directions of their electric fields using different "windows", or, in other words, specific regions of the detector. of the SE plume on the detector **b**. A central window **c** results in a drop in electron counts at the edges of the p-doped square where the SE plume is deflected out of the window used for producing the micrograph. This window was physically constructed using electrical tape as a spatial mask on the glass capillary array that back the phosphor screen **c** (inset). Windows on the left **d** or right **e** sides of the plume show enhanced sensitivity for SE emission from the left and right sides of the p-doped square, leveraging the lateral deflection of the SE plume. Spatial masks on the phosphor used to acquire these micrographs are provided in Fig. S9. Conversely, windows at the bottom or top of the plume can selectively image the top **f** or bottom **g** sides of the p-doped square. These micrographs were acquired using the spatial mask on the phosphor in **c** inset, instead using the sample electrode voltage to move the SE plume on the screen to select specific features. Specific sample electrode voltages used for each micrograph are provided in relevant panels **c**, **f**, **g**. The same structure was used to record all micrographs, with white arrows in **d**-**g** highlighting the side of the square that is selectively imaged. In all micrographs, darker blue denotes higher counts, albeit contrast is not directly comparable between micrographs as different voltages on the MCP/phosphor were used to optimize each. Colors and letters in the schematic in **b** correspond to their respective micrographs in **c**-**g**. Scale bar in **c** 50 μm.

To this end, we selectively imaged features on our planar p-n junction sample based on their electric field vectors (Fig. 3). To do so, different portions of the SE plume on the viewing screen were used to record micrographs (Fig. 3b). Micrographs were recorded from selected areas on our detector by spatially masking the viewing screen with electrical tape, thereby only selecting electrons with specific trajectories (Fig. 3c, inset). When this spatial mask was in the center of the SE plume, and thus only capturing electrons with nearly vertical emission angles, micrographs were recorded with the low-count halo at the p-n interfaces (Fig. 3c), as the interfacial electric fields deflected the SE plume away from the window seen by the PMT resulting in the decreased counts. Conversely, this window could be moved off-center to selectively image the sides of the p-doped square based on the direction of the interfacial electric fields. These windows used are shown schematically in Fig. 3b. A window on the left or right of the plume resulted in selective imaging of the left (Fig. 3d) or right (Fig. 3e) sides of the square. As previously noted, the SEs are deflected laterally away from the p-doped material exact at the p-n interfaces and do not experience substantial lateral deflections far from the interfaces, resulting in this selective imaging of the left or right sides of the p-doped square with left or right windows. Only SEs generated very near the interfaces experience deflection into these off-center windows. Conversely, windows on the bottom or top of the plume selectively imaged the top (Fig. 3f) or bottom (Fig. 3g) sides of the square. Left and right sides (Fig. 3d,e) were recorded with spatial masks with windows on the left and right sides of the phosphor screen, respectively (Fig. S9). Top and bottom (Fig. 3f,g) were recorded with a central window (Fig. 3c,inset), but using the sample electrode voltage to move the SE plume vertically on the detector to select the desired features. SE trajectory simulations (Fig. S10), demonstrate that the SEs deflected away from the p-doped region on the top (bottom) side impact the detector at lower (higher) position on the phosphor screen, explaining the sample electrode voltage dependence observed in these experiments (Fig. 3b,f,g). Effectively, SEs with initial velocities toward the repeller electrode encounter a steeper voltage gradient upon exiting the shroud and more quickly get turned toward the viewing screen than SEs with only vertical velocity, thus enhancing the lower part of the plume (Fig. 3b, green window). Conversely, SEs with initial velocities away for the repeller electrode encounter a shallower voltage gradient and more slowly get turned compared to SEs with only vertical velocity and enhance the upper part of the plume (Fig. 3b, yellow window). These experiments, in effect, demonstrate the ability to directly map electric fields using momentum-resolved SEs.

**Resolving Buried Interfaces**

Following the observations of the built-in field contrasts, we investigated the effects of buried interfaces on the surface, since such effects are important in non-destructive probing of buried electronics. Remarkably, the electric field contrast at the p-n interfaces can be observed even through a 50 nm thick sputtered oxide layer (Fig. 4a,b). At this thickness, the SEs recorded are generated from the $SiO_2$ rather than the underlying p-n junction material.[27, 28] The edge contrast at the sides of the p-doped square are still present (Fig. 4a, '$SiO_2$ coated') as they were for the pristine sample (Fig. 4a, 'uncoated') which only has a thin (3.5 nm) thermal oxide layer. This surprising effect can be explained by surface band bending in the oxide overlayer due by the buried interface, adding on to the growing literature on how deep buried structures impact surface electronic states in devices.[20, 29, 30] Studying these dynamics requires a technique that is sufficiently surface-sensitive to resolve these surface dynamics, but is also not constrained by surface defects that pin the Fermi level. SEs, which are generated from a finite depth of several nanometers below the surface, are sensitive to near-surface electric fields but are not pinned by states exactly at the surface. Furthermore, the observation of electric field contrast through the oxide layer demonstrates that we can observe electronic effects above a p-n interface and showcases a capability for non-destructive probing of structures under a passivating layer, of significance for engineering of layered devices or devices with air-sensitive materials.

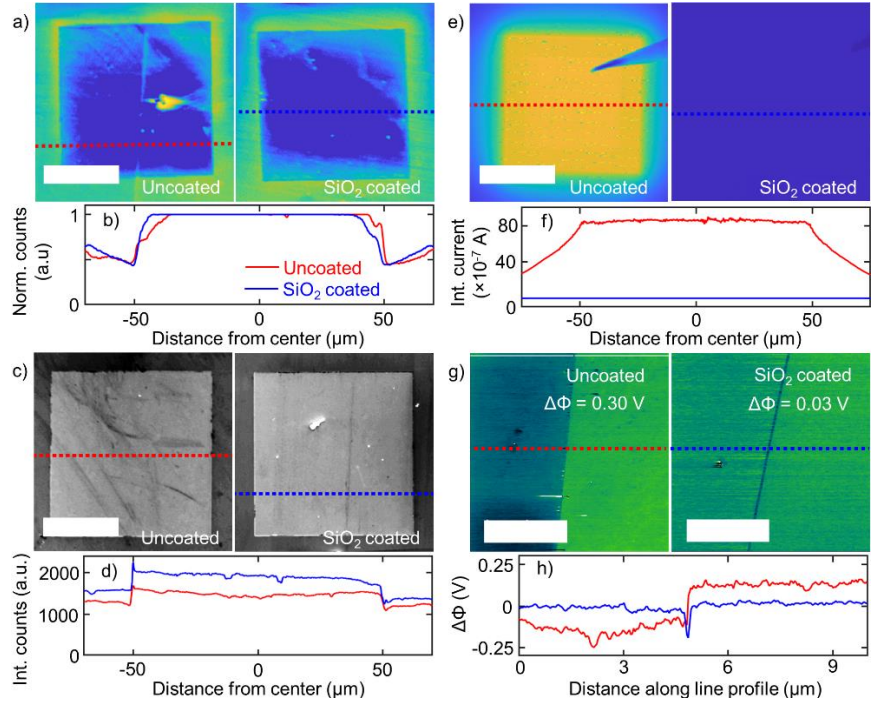

**Figure 4| Comparison between SE plume imaging and complementary techniques. a** Micrographs of a p-doped square, both pristine and with 50 nm of sputtered SiO$_2$ recorded using our detector, using spatial mask at the center of the SE plume. Edge contrast due to lateral electric fields are observed in both micrographs, and in line profiles **b** taken across the p-doped squares, with very similar magnitudes for both samples. This edge contrast was not observed in ETD **c** micrographs or **d** line profiles of these samples. Edge features in the line profiles **d** are likely topographical features (Fig. S6). **e** The lateral electric fields across a p-n junction can also be observed by EBIC, in which a micromanipulator probe is used to measure current generated by the electron beam, but this requires direct electrical contact, and thus cannot be used to probe fields under the sputtered oxide layer ('SiO$_2$ coated'). **f** Line profiles across EBIC micrographs. **g** Similarly, KPFM, which is used to measure work functions of surfaces, can resolve the differences in fermi level of the pristine sample ($\Delta\Phi \sim 0.3$ V). However, like EBIC, KPFM cannot probe dynamics beneath the sputtered oxide. **h** Line profiles across KPFM micrographs. In all micrographs, red or blue dotted lines denote where line profiles were taken, with colors corresponding to their respective line profiles; red for the uncoated sample and blue for the sputter coated sample. Scale bars 50 μm in **a**, **c**, **e**, 4 μm in **g**. Norm. counts; Normalized counts. Int. counts; integrated counts. Int. current; integrated current. $\Delta\Phi$; change in work function.

This capability is unique to our detector. As a comparison, p-doped squares, both uncoated and SiO$_2$ coated, were imaged with complimentary techniques; SEM imaging using an ETD (Fig. 4c), electron beam induced current (EBIC) measurements (Fig. 4e), in which a micromanipulator probe is used to collect current generated in a sample by an SEM electron beam, and kelvin probe force microscopy (KPFM), an atomic force microscopy technique capable of measuring surface work functions (Fig. 4g).

Unsurprisingly, the electric field contrast was not observed in an ETD micrograph of the sample under the sputtered oxide layer (Fig. 4c, 'SiO$_2$ coated'), as it also was not observed in the uncoated sample (Fig. 4c,

'uncoated'). However, the doping contrast was observed in both micrographs, as it was for micrographs recorded using our detector (Fig. 4a).

EBIC is capable of measuring electric fields across a p-n junction, with the electron beam producing electron-hole pairs in a sample that can diffuse across the interface for and be collected by the probe.[31, 32] This measurement resulted in the bright halo around the p-doped square in the micrograph of the uncoated p-doped square (Fig. 4e, 'uncoated'). However, it requires electrical contact between the sample and probe and can only measure sub-surface electric fields. Therefore, EBIC is precluded from measuring buried structures. Consequently, there was no EBIC signal through the thick $SiO_2$ layer of the coated sample (Fig. 4e, '$SiO_2$ coated') due to the poor conductivity of the oxide material. EBIC contrast can be recovered by pressing the probe through the oxide layer (Fig. S6a), but this process is necessarily invasive and destructive. Thus, the ability to resolve electric field effects through a passivating layer is unique to momentum-resolved SE imaging.

KPFM can be used for measuring surface potentials in a material,[33] and is an exceptional tool to characterize p-n junctions.[34] A work function difference of ~0.3 V was measured across the p-n junction of our sample (Fig 4g, 'uncoated'). However, since KPFM is sensitive to the surface contact potential difference, it cannot measure a difference in work function when the sample is sputter-coated with 50 nm $SiO_2$ (Fig. 4g, '$SiO_2$ coated').[35, 36] Since SEs are generated from several nm below the surface, this is not an issue for our detector.

Lastly, other techniques can indirectly study electric fields. Scanning ultrafast electron microscopy (SUEM)[20, 25, 37, 38] can image the effects of electric fields and the resulting charge carrier motion, but with much more rigorous data collection and analysis and with heavier SEM modifications. Alternately, secondary ion mass spectroscopy (SIMS)[39] and scanning capacitance microscopy (SCM)[40, 41] can provide information about dopants and charge carriers, but cannot directly measure electric fields. Overall, the ability to directly image electric fields, both above and below a surface and even through passivating oxide layers, using momentum-resolved SEs is a unique capability of this detector that can be applied in imaging interfacial dynamics in a variety of semiconductor devices and magnetic samples.

**Materials analysis with energy resolved electrons**

Simulations of electron trajectories through our detector (Fig. S2, S3) demonstrate the ability to resolve electrons by energies along the z-axis of our detector, thereby enabling quasi-SE energy spectroscopy for measuring surface potentials.[9, 12] As described earlier, SEs with equivalent energies moving towards or away from the repeller electrode will impact the viewing screen at different heights due to the different effective voltage gradients they encounter upon exiting the shroud. As a result, the energy spectrum is blurred and requires deconvolution for quantitative spectral analysis. Despite this blurring, qualitative comparisons are possible. To verify this ability to learn about surface potentials despite this velocity dependent blurring, the full SE plume of an unpatterned p-type Si wafer was recorded (Fig. 5a-d), enabling extraction of SE energy and momentum spectra. This process was also repeated for isolated p- and n-doped regions of the p-n junction sample to reveal differences in their SE energy spectra (Fig. 5e,f) and highlight possibility of semi-quantitative materials analysis with energy-resolved SEs.

We performed demonstrations of energy- and momentum-resolved SE spectroscopy with our detector using an unpatterned p-type Si wafer (Fig. 5 a-d). Changing the sample electrode voltage offers a way to control the position of the SE plume on the viewing screen (Fig. 5a, S3,S4), allowing calibration of the energy axis and ensuring homogeneity of signal across the energy spectra by mitigating edge effects or other irregularities that vary by position on the viewing screen. Simulations (Fig. S4) suggest that there is a 1 eV: 1 V correspondence between SE energy and sample voltage when scanning the sample electrode voltage and collecting signal from the same position on the viewing screen for each voltage. Thus, a 3D plot of the full SE plume (Fig. 5b) was obtained by scanning the sample electrode voltage from -205 to -125 V. Data processing details are provided in the supporting information (Fig. S11) From this data set, we estimate the energy resolution of our detector to be ~70 meV/pixel at the chosen experimental settings, on account of a roughly 140 pixel shift in plume height with a 10 V change in sample voltage (Fig. 5a) and an energy range of ~45 eV from ~640 pixels diameter of the screen. These parameters can be changed by altering electron optics parameters, incorporating lensing optics, or by simply using a higher resolution digital camera. Further, the specific electron energy range can be tuned with the sample electrode voltage for various applications—for instance, changing the sample electrode voltage by ~ +100 V may enable Auger electron detection.

Energy (Fig. 5c) and momentum (Fig. 5d) spectra were extracted from the 3D plume. Details on this analysis are provided in the supporting information. Due to the above described convolution between the angular spreads and impact heights on the viewing screen of the generated SEs, there is a distortion in shape between prior published SE spectra,[11] which exhibit the rising edge of a peak within a few eV and a long tail at higher energies, and our raw (i.e. not deconvolved) recorded SE energy spectrum, which is very symmetric and has ~54 V of full-width-tenth-max (FWTM). SIMION simulations of electron trajectories through our detector using a published SE energy spectrum[11] and various angular distributions (Fig. S12-S14) were performed to extract angular information of the SE plume from our recorded spectrum. Treating the angular spread as a 40° (half angle) cone oriented along the +$z$ axis results in good qualitative agreement between simulations results (Fig. 5c, grey histogram) and our recorded energy spectrum (Fig. 5c, red curve). A peak in the histogram at around 20 graphics units is due to the electrons clipping on the slit of the shroud electrode, which was observed experimentally, particularly when data was collected from locations away from the center of the slit along its short axis. This, combined with analysis of peak broadening with angular spread (Fig. S12) suggest that the SE distribution of our sample is well represented as a ~30 – 40° (half angle) cone with SEs generated at a shallow angle (i.e. nearly normal to the plane of the sample) contributing to the center of the height distribution of the projected SE plume on the detector and SEs generated at wide angles contributing to the extremities of the height distribution of the projected SE plume on the detector (Fig. S14), but moreover demonstrates the ability to extract quantitative information about SE distributions from the recorded SE plume with sufficient modeling. Thus, from the SE plume, one can feasibly back-calculate surface electronic band structures or densities of states, akin to ARPES[5] but instead using an SEM. Notably, however, SEs experience sub-surface inelastic scattering[2] processes, such as electron-electron scattering, before escaping the sample that broaden their energy distributions and will make calculations of band structures challenging even with deconvolved SE energy spectra.

The SE momentum spectrum (Fig. 5d) is asymmetric (asymmetry in the plume in the plane parallel to the surface), likely reflecting slight asymmetries in the current detector design, as well as effects from stray electromagnetic fields that distort the SE plume. The sample is an unpatterned Si wafer, so there are no topographical or interfacial features present that would explain this asymmetry, nor can it be explained by its cubic crystal structure. Instead, it most likely arises from external electromagnetic fields that impact the shape of the SE plume as it travels to the viewing screen, as well as slight misalignments of the detector. For example, in the current iteration, it is likely that the repeller electrode is not perfectly parallel to the axis of the detector screen. Future iterations of the detector will include more shielding to mitigate the effects of electromagnetic fields and be more rigid and robust to address this shortcoming.

Comparisons of the recorded SE energy spectra for p- and n- doped Si (Fig. 5e,f) are promising towards future applications in quantitative materials analysis utilizing this detector. Energy spectra acquired from p- and n- doped regions on the p-n junction sample display several differences that reflect their underlying compositional and band structure differences. The spectra were acquired from featureless, isolated areas away from interfaces to mitigate the effects of the lateral electric fields. Fewer electrons were generated from the n-doped region than the p-doped region, reflecting the doping contrast of these materials.[21-23, 25] Further, the peak energy is higher (less negative sample electrode voltage) for the SE electrons generated from the n-doped region, likely due to its higher fermi level. Parameters from fitting these peaks to a mixed Gaussian/Lorentzian peak shape are provided in Table S2. These differences are promising for quantitative materials analysis by exploiting the energy resolution of our detector with proper calibration and deconvolution.

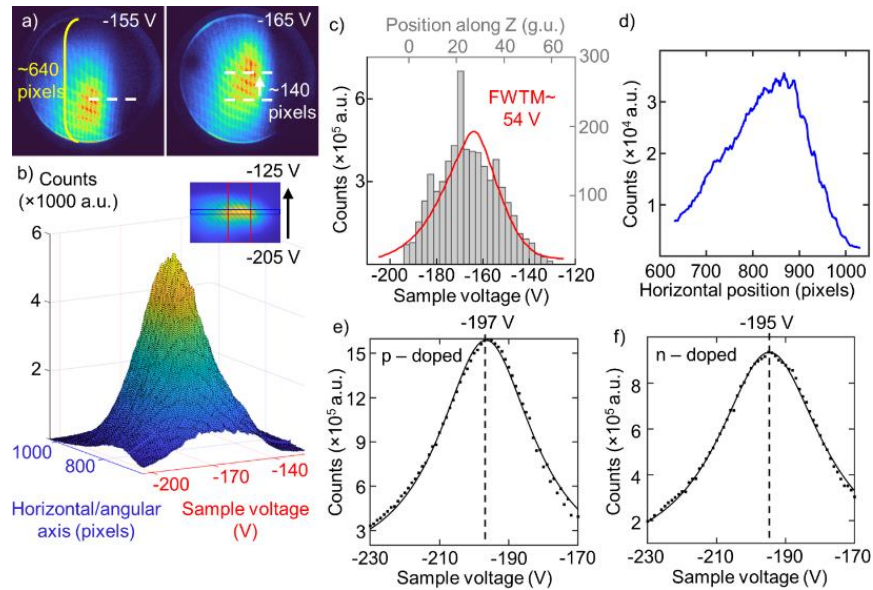

**Figure 5| SE plume from Si. a** Pictures of the SE plume from an unpatterned p-type Si wafer taken at -155 and -165 V on the sample electrode. The change in sample electrode voltage shifts the position of the SE plume vertically on the phosphor screen, which is applied for energy spectroscopy. The grid pattern on these images is a projection of the Cu mesh at the front of the mu metal tube (Fig. S11). **b** A 3D plot of the SE plume from this sample, counts versus horizontal position on the camera versus sample electrode voltage. Data processing details are provided in the supporting information (Fig. S12). Inset: top-down color plot of the SE plume. Striations are an artifact from the cu mesh. Red and blue rectangles were integrated along their horizontal and sample electrode voltage axes, respectively, to produce **c** an SE energy spectrum (red curve) and **d** an SE momentum spectrum. The SE energy spectrum is overlain on a histogram (grey axes and chart) of electron impact heights from electron trajectory simulations of our detector, effectively an energy spectrum, using a published SE spectrum[11] and a 40° conical velocity distribution (Fig. S12-S14). Qualitative agreement between simulation results and experimental results suggest that the electron angular distribution is well represented by a 30 – 40° cone and highlight the capability for extracting quantitative information from an SE distribution with our detector. SE energy spectra were also collected from **e** p- and **f** n-doped Si regions on the planar p-n junction sample and fit to mixed Gaussian/Lorentzian peak shapes. These spectra differ in ways consistent with expectations from their dopant profiles. Specifically, the smaller peak size and more positive peak voltage of the n-doped material reflects the expected dopant contrast and higher average SE energies due to the higher fermi level for n-doped Si compared to p-doped Si. a.u.; arbitrary units. g.u.; graphics unit, or 1 mm at simulation scale.

## Conclusions

Despite ubiquity in the field of materials science, SEM, as it is routinely applied, misses out on the information stored within SEs that can unveil hidden information about the electronic structure of a sample surface. By simple modifications to a standard SEM, we were able to extract momentum- and energy-information contained within SE distributions by direct imaging of the SE plume. Exploiting our momentum resolution, we were able to map interfacial electric fields across a p-n junction even when buried beyond depth accessible by SEM and by exploiting energy resolution we were able to identify differences in SE energy distributions for differently doped samples that reflect their different chemistries. These results show that SEM plume imaging can be used to provide precise physical meaning to SEM image contrast for in depth probing of material structures in a non-invasive manner, with notable applications in studying interfacial structures in passivated semiconductor devices, mapping electromagnetic fields, or likely even quantitative band structure measurements in a relatively simple way.

**Acknowledgements:**

This paper describes objective technical results and analyses. Any subjective views or opinions that might be expressed in the paper do not necessarily represent the views of the US Department of Energy (DOE) or the US government.

**Funding:**

This work was mainly supported by the U.S. Department of Energy, Office of Science, Office of Basic Energy Sciences, under Grant DE-SC0021070 (F.M.A., A.A.T., and S.K); and by the Division of Chemical Sciences, Geosciences and Biosciences, Office of Basic Energy Sciences (BES) funded by the US DOE (E.J.S., and D.C.); and by the Laboratory Directed R&D program of Sandia National Laboratories (C.P. and C.Y.N.).

**Author Contributions:**

D.W.C. conceived of the original detector design in 2015, with the first iteration being tested and designed by E.J.S. Starting in 2021 C.P. and E.J.S. iterated on the design to produce the current design which was used to acquire the data in this manuscript. Starting in 2023, F.M.A and D.W.C. ran SIMION simulations of this design for guiding usage and developing next iterations, with the assistance of E.J.S. and C.P. Samples were provided by C.Y.N. and A.A.T. F.M.A, A.A.T., D.C. and S.K. conceived of experiments performed in this manuscript and aided in analysis of results. Sputter coating of a sample was performed by F.N. KPFM measurements of samples was performed by L.H. under the supervision of A.J.M. The original draft was written primarily by F.M.A., D.W.C, and S.K., with all authors performing editing. D.W.C. and S.K. supervised the project.

**Data and materials availability**: All data needed to evaluate the conclusions in the paper are present in the paper and/or the Supplementary Materials.

References:

(1) Everhart, T. E.; Thornley, R. F. M. Wide-band detector for micro-microampere low energy electron currents. *J. Sci. Instrum.* **1960**, *37*, 246-248. DOI: 10.1088/0950-7671/37/7/307.
(2) Seiler, H. Secondary Electron Emission in the Scanning Electron Microscope. *J. Appl. Phys.* **1983**, *54*, R1-R18. DOI: https://doi.org/10.1063/1.332840.

(3) Li, J.; Malis, T.; Dionne, S. Recent advances in FIB–TEM specimen preparation techniques. *Materials Characterization* **2006**, *57*, 64-70.

(4) Wirth, R. Focused Ion Beam (FIB) combined with SEM and TEM: Advanced analytical tools for studies of chemical composition, microstructure and crystal structure in geomaterials on a nanometre scale. *Chemical Geology* **2009**, *261*, 217-229.

(5) Damascelli, A. Probing the Electronic Structure of Complex Sysesms by ARPES. *Phys. Scr.* **2004**, *2004*, 61-74. DOI: 10.1238/Physica.Topical.109a00061.

(6) Richards, B. P.; Footner, P. K. Failure analysis in semiconducto devices - rationale, methodology and practice. *Microelectronics Journal* **1984**, *15* (1), 5-25. DOI: https://doi.org/10.1016/S0026-2692(84)80003-8.

(7) Chin, C.; Kim, K.; Kim, J.; Ko, W.; Yang, Y.; Lee, S.; Jun, C. S.; Kim, Y. S. Fast, exact and non-destructive diagnoses of contact failures in nano-scale semiconductor device using conductive AFM *Sci. Rep.* **2013**, *3*, 2088. DOI: https://doi.org/10.1038/srep02088.

(8) Lampman, S.; Mulherin, M.; Shipley, R. Nondestructive testing in failure analysis. *J. Fail. Anal. Prev.* **2022**, *22*, 66-97. DOI: ttps://doi.org/10.1007/s11668-021-01325-1.

(9) Chung, M. S.; Everhart, T. E. Simple calculation of energy distribution of low-energy secondary electrons emitted from metals under electron bombardment. *J. Appl. Phys.* **1974**, *45*, 707-709. DOI: 10.1063/1.1663306.

(10) Han, W.; Zheng, M.; Banerjee, A.; Luo, Y. Z.; Khursheed, A. Quantitative material analysis using secondary electron energy spectromicroscopy. *Sci. Rep.* **2020**, *10*, 22144. DOI: 10.1038/s41598-020-78973-0.

(11) Srinivasan, A.; Han, W.; Zheng, M.; Khursheed, A. Characterization of Materials Using the Secondary Electron Energy Spectromicroscopy Technique. *Opt. Mater.: X* **2021**, *12*, 100121. DOI: 10.1016/j.omx.2021.100121.

(12) Zhou, Y.; Fox, D. S.; Maguire, P.; O'Connell, R.; Masters, R.; Rodenburg, C.; Wu, H.; Dapot, M.; Chen, Y.; Zhang, H. Quantitative Secondary Electron Imaging for Work Function Extraction at Atomic Level and Layer Identification of Graphene. *Sci. Rep.* **2016**, *6*, 21045. DOI: https://doi.org/10.1038/srep21045.

(13) Appelt, G. Fine Structure Measurements in the Energy Angular Distribution of Secondary Electrons from a (110) Face of Copper. *Phys. Status Solidi B* **1968**, *27*, 657-669.

(14) Best, P. E. Energy- and Angular-Dependent Secondary-Electron Emission from a Silicon (111) 7 X 7 Surface. Emission from Bulk States. *Phys. Rev. B* **1976**, *14*, 606-619. DOI: https://doi.org/10.1103/PhysRevB.14.606.

(15) Jones, G. A. Magnetic contrast in the scanning electron microscopy: An appraisal of techniques and their applications. *Journal of Magnetism and Magnetic Materials* **1978**, *8*, 263-285. DOI: https://doi.org/10.1016/0304-8853(78)90096-3.

(16) Akamine, H.; Okumura, S.; Farjami, S.; Murakami, Y.; Nishida, M. Imaging of surface spin textures on bulk crystals by scanning electron microscopy. *Sci. Rep.* **2016**, *6*, 37265. DOI: https://doi.org/10.1038/srep37265.

(17) Chang, C. C. Auger Electron Spectroscopy. *Surf. Sci.* **1971**, *25*, 53-79. DOI: 10.1016/0039-6028(71)90210-X.

(18) Wilkinson, A. J.; Britton, T. B. Strains, Planes, EBSD in Materials Science. *Mater. Today* **2012**, *15*, 366-376. DOI: 10.1016/S1369-7021(12)70163-3.

(19) Nowell, M. M.; Wright, S. I. Phase differentiation via combined EBSD and XEDS. *J. Microsc.* **2004**, *213*, 296-305. DOI: 10.1111/j.0022-2720.2004.01299.x.

(20) Ellis, S. R.; Bartelt, N. C.; Léonard, F.; Celio, K. C.; Fuller, E. J.; Hughart, D. R.; Garland, D.; Marinella, M. J.; Michael, J. R.; Chandler, D. W.; et al. Scanning Ultrafast Electron Microscopy Reveals

Photovotage Dynamics at a Deeply Buried p-Si/SiO$_2$ interface. *Phys. Rev. B* **2021**, *104*, l161303. DOI: 10.1103/PhysRevB.104.L161303.
(21) Frank, L.; Hovorka, M.; El-Gomati, M. M.; Müllerová, I.; Mika, F.; Mikmeková, E. Acquisition of the Dopant Contrast in Semiconductors with Slow Electrons. *Journal of Electron Spectroscopy and Related Phenomena* **2020**, *241*, 146836. DOI: https://doi.org/10.1016/j.elspec.2019.03.004.
(22) Chee, K. W. A.; Rodenburg, C.; Humphreys, C. J. High resolution dopant profiling the SEM, image widths and surface band bending. *Journal of Physics: Conference Series* **2008**, *126*, 012033. DOI: doi:10.1088/1742-6596/126/1/012033.
(23) Chee, A. K. W. Enhancing doping contrast and optimising quantification in the scanning electron microscope by surface treatment and Fermi level pinning. *Sci. Rep.* **2018**, *8*, 5247. DOI: https://doi.org/10.1038/s41598-018-22909-2.
(24) Sekiguchi, T.; Kimura, T.; Iwai, H. SEM observation of p-n junction in semiconductors using fountain secondary electron detector. *Superlattices and microstructures* **2016**, *99*, 165-168. DOI: https://doi.org/10.1016/j.spmi.2016.03.020.
(25) Perez, C.; Ellis, S. R.; Alcorn, F. M.; Smoll, E. J.; Leonard, F.; Chandler, D. W.; Talin, A. A.; Bisht, R. S.; Ramanathan, S.; Goodson, K. E.; et al. Picosecond carrier dynamics in InAs and GaAs revealed by ultrafast electron microscopy. *Science Advances* **2024**, *10*, eadn8980. DOI: https://doi.org/10.1126/sciadv.adn8980.
(26) Anderson, M. L.; Nakakura, C. Y.; Kellogg, G. L. *Imaging doped silicon test structures using low energy electron microscopy*; SAND2009-7981; OSTI.gov, 2010. https://www.osti.gov/servlets/purl/993607.
(27) McFeely, F. R.; Cartier, E.; Yarmoff, J. A.; Joyce, S. A. Low-energy-electron escape lengths in SiO$_2$. *Phys. Rev. B* **1990**, *42* (8), 5191-5200. DOI: https://doi.org/10.1103/PhysRevB.42.5191.
(28) Yi, W.; Jeong, T.; Yu, S.; Lee, J.; Jin, S.; Heo, J.; Kim, J. M. Study of the secondary-electron emission from thermally grown SiO$_2$ films on Si. *Thin Solid Films* **2001**, *397* (1-2), 170-175. DOI: https://doi.org/10.1016/S0040-6090(01)01492-4.
(29) Çopuroğlu, M.; Sezen, H.; Opila, R. L.; Suzer, S. Band-bending at buried SiO$_2$/Si interface as probed by XPS. *ACS Appl. Mater. Interfaces* **2013**, *5*, 5875-5881. DOI: https://doi.org/10.1021/am401696e.
(30) Ishioka, K.; Brixius, K.; Beyer, A.; Rustagi, A.; Stanton, C. J.; Stolz, W.; Volz, K.; Hofer, U.; Petek, H. Coherent phonon spectroscopy characterizatino of electronic bands at buried semiconductor heterointerfaces. *Appl. Phys. Lett* **2016**, *108*. DOI: https://doi.org/10.1063/1.4941397.
(31) Lahreche, A.; Babichev, A. V.; Beggah, Y.; Tchernycheva, M. Modeling of the electron beam induced current signal in nanowires with an axial p-n junction. *Nanotechnology* **2022**, *33*, 395701. DOI: DOI 10.1088/1361-6528/ac7887.
(32) Haney, P. M.; Yoon, H. P.; Gaury, B.; Zhitenev, N. B. Depletion region surface effects in electron beam induced current measurements. *J. Appl. Phys.* **2016**, *120*, 095702. DOI: https://doi.org/10.1063/1.4962016.
(33) Nakamura, T.; Ishida, N.; Sagisaka, K.; Koide, Y. Surfface potential imaging and characterization of a GaN p-n junction with Kelvin probe force microscopy. *AIP Advances* **2020**, *10*, 085010. DOI: https://doi.org/10.1063/5.0007524.
(34) S. Saaf; Rosenwaks, Y. Local measurement of semiconductor band bending and surface charge using Kelvin probe force microscopy. *Surf. Sci.* **2005**, *574*, L35-L39.
(35) Baumgart, C.; Helm, M.; Schmidt, H. Quantitative dopant profiling in semiconductors: A Kelvin probe force microscopy model. *Phys. Rev. B* **2009**, *80*, 085305.
(36) Melitz, W.; Shen, J.; Kummel, A. C.; Lee, S. Kelvin probe force microscopy and its applications. *Surf. Sci. Rep.* **2011**, *66*, 1-27.

(37) Najafi, E.; Ivanov, V.; Zewail, A.; Bernardi, M. Super-diffusion of excited carriers in semiconductors. *Nat. Commun.* **2017**, *8*, 15177.

(38) Najafi, E.; Scarborough, T. D.; Tang, J.; Zewail, A. Four-dimensional imaging of carrier interface dynamiucs in p-n junctions. *Science* **2015**, *2015*, 164. DOI: https://doi.org/10.1126/science.aaa0217.

(39) Eswara, S.; Pshenova, A.; Lentzen, E.; Nogay, G.; Lehmann, M.; Ingenito, A.; Jeangros, Q.; Haug, F.-J.; Valle, N.; Philipp, P.; et al. Secondary ion mass spectrometry quantification of boron distribution in an array of silicon nanowires. *MRS Commun* **2019**, *9*, 916-923. DOI: https://doi.org/10.1557/mrc.2019.89.

(40) Zavyalov, V. V.; McMurray, J. S.; Williams, C. C. Advances in experimental technique for quantitative two-dimensional dopant profiling by scanning capacitance microscopy. *Rev. Sci. Instrum.* **1999**, *70*, 158-164. DOI: https://doi.org/10.1063/1.1149558.

(41) Kimura, K.; Kobayashi, K.; Yamada, H.; Matsushige, K.; Usuda, K. Scanning capacitance force microscopy imaging of metal-oxide-semiconductor field effect transistors. *J. Vac. Sci. Technol. B* **2005**, *23*, 1454-1458. DOI: https://doi.org/10.1116/1.1941188.

# Supporting information for Resolving the Electron Plume within a Scanning Electron Microscope

Francis M. Alcorn*[1], Christopher Perez[1,2], Eric J. Smoll[1], Lauren Hoang[3], Frederick Nitta[3,4], Andrew J. Mannix[4], A. Alec Talin[1], Craig Y. Nakakura[5], David W. Chandler*[1], Suhas Kumar*[1]

[1] *Sandia National Laboratories, Livermore, CA, USA*
[2] *Department of Mechanical Engineering, Stanford University, Stanford, CA, USA*
[3] *Department of Electrical Engineering, Stanford University, Stanford, CA, USA*
[4] *Department of Materials Science and Engineering, Stanford University, Stanford, CA, USA*
[5] *Sandia National Laboratories, Albuquerque, NM, USA*

Section S1: SEM settings

The detector was built onto an FEI XL30 Schottky field emission gun (SFEG) scanning electron microscope (SEM), using an existing port on the sidewall of the microscope. All experiments were performed in spot size 5 with an accelerating voltage of 10 kV and a 200 μm condenser aperture.

Section S2: Detector design

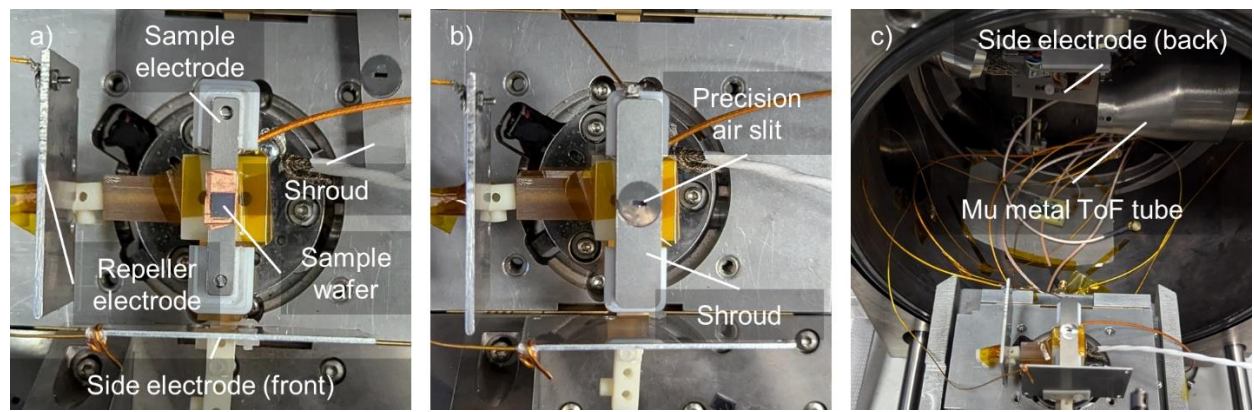

**Figure S1| Photographs of electron optics.** (a) Electron optics assembly without the shroud electrode in place to show the sample wafer on the sample electrode. (b) Fully assembled electron optics assembly with shroud covering sample electrode. A precision air slit on the shroud electrode allows the primary electron beam to reach the sample and the generated SEs to escape the shroud. (c) View inside sample chamber, highlighting the mu metal ToF tube which leads to the MCPs/phosphor screen, and the grounded side electrode built into the sample chamber. In all pictures, SEs travel towards the right to impact the detector screen.

Electrostatic electron optics are used to project the secondary electron (SE) plume onto the viewing screen. These electrodes are: repeller electrode, front side electrode, back side electrode, sample electrode, shroud, and a grounded me metal time-of-flight (ToF) tube. These electrodes were made with machined stainless steel. Shape and dimensions for each are provided in Table S1. Electrodes were mounted onto the stage using a custom scaffold made of macor and PFTE, aside from the back electrode which was mounted onto the sample chamber wall with copper wire. The sample electrode sits at the top of the central macor piece and has a PTFE boat/scaffold designed on it for placing the shroud electrode, a hollow rectangular shell, fully around it. This fully encloses the sample and blocks electromagnetic interference, particularly from the repeller electrode, just as SEs are generated at the sample surface. These SEs are accelerated out from the shroud electrode through a slit on the top of the shroud. The slit is a precision air slit (Edmund Optics), which is a thin, circular stainless-steel foil 9.5 mm in diameter with a slit of varying geometry in the center. This is placed within a 1 cm circular cutout centered on top of the shroud. Segmentation of the slit and shroud afford an additional degree of freedom in slit geometry (size, shape orientation) that can be used for controlling SE plume behavior as it appears on the viewing screen. Experiments performed in this study used a 3 by 0.2 or 0.5 mm rectangular slit.

**Table S1.** Electrode shapes and dimensions.

| Electrode | Shape and dimensions |
|---|---|
| Repeller | Rectangular plate; 57 mm × 31 mm × 1 mm |
| Front | Rectangular plate; 54 mm × 33 mm × 1 mm |
| Back | Rectangular plate; 54 mm × 33 mm × 1 mm |
| Sample | Rectangular prism; 39 mm × 5 mm × 5 mm |
| Shroud | Rectangular shell; 43 mm × 10 mm × 10 mm. Wall width 1 mm |

All electrodes were connected to high voltage power supplies (Stanford Research Systems Model PS325 or 350) via electrical feedthroughs into the SEM, aside from the back electrode which was grounded to the sample chamber walls. The sample electrode was biased to roughly -150 to -200 V to push the SE plume out of the shroud which was kept grounded. The repeller electrode was biased at -3000 to -5000 V to project this plume onto the viewing screen. The front electrode was biased at up to ±500 V to center the SE plume laterally on the viewing screen.

Once pushed towards the viewing screen by the repeller electrode, SEs enters a grounded mu-metal ToF tube through a Cu mesh. The SEs continue along the length of the mu metal ToF tube and impact the viewing screen, a 75 mm Photonis Advanced performance detector composed of two microchannel plates (MCPs) (12/10 pitch/pore size, 8° bias angle, 60:1 aspect ratio) and a P47 phosphor. The MCPs were biased at ~900 V each (~1800 V total, divided in half) and the phosphor up to 4000 V. Voltages were supplied by high voltage power supplies (Stanford Research Systems Model PS350). The MCPs act as a gain medium to amplify the incident electrons and the phosphor converts these amplified electrons into photons for collection and digitization. Voltages on the repeller electrodes were ~ -3000 – 4000 V, and voltages of up to several hundred volts on the side electrodes were used to laterally align the SE plume.

For data collection and digitization, a sufficient photodetector for each experiment was used to collect signal from the phosphor screen. For recording SEM micrographs, a photomultiplier tube (PMT) was used as a fast photodetector, synchronized to the scan of the electron beam with custom LabView code. To record images of interfacial electric fields, the PMT only collected data from a window on the viewing screen by spatially filtering the viewing screen (Fig. 3c inset, S10). The PMT was biased to 900 V by a Bertan Model 205A-05R High Voltage Power Supply and interfaced to a computer via a National Instruments USB-6363 multifunction DAQ. In other experiments, pictures of the electron plume on the phosphor were taken with a digital camera (Basler acA 1920-155 µm equipped with a Cosmicar television lens, 0.1-1 s exposure time).

Section S3: SIMION 8.1.1.32 electrodynamics simulations

SIMION simulations were performed of electrons within a scale model of our detector to validate the design (Fig. S2-S4). In these simulations, electrons were generated just above the sample surface with velocities in a 'filled in' conical distribution (half angle 2°) centered along the +z direction. In all simulations, repeller voltages were -3500 V, sample voltages were -240 V unless noted otherwise, and all other electrode voltages were 0. These simulations demonstrate the ability to separate electrons by energies in the z-dimension (Fig. S2) and that the sample voltage can be used to control the impact height of electrons on the detector (Fig. S3, S4), exploited for performing SE energy spectroscopy (Fig. 5).

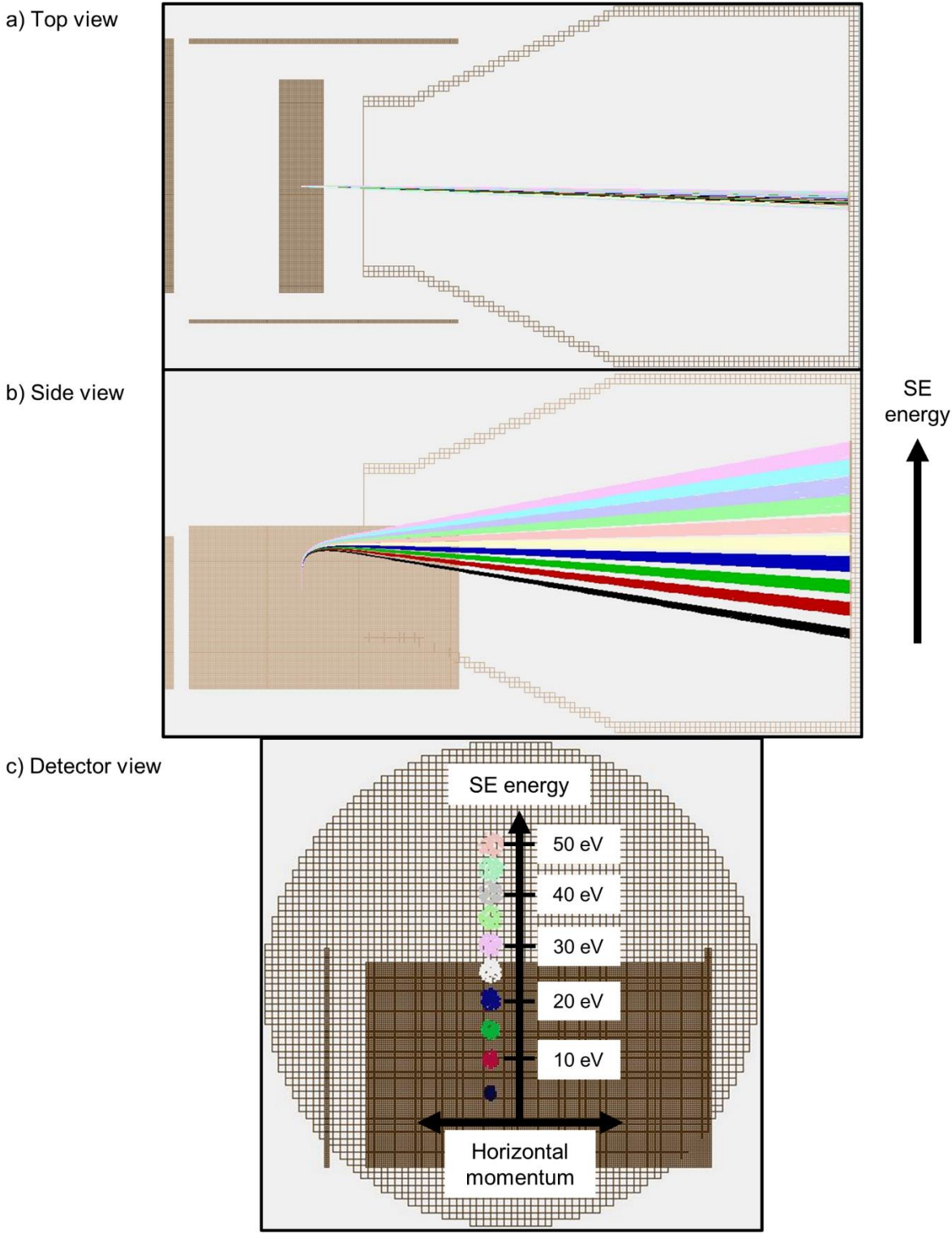

**Figure S2| Simulating trajectories of different energy (5 – 50 eV) electrons through our detector.** (a) Top view, (b) side view, and (c) view of the detector face. (b and c) Higher energy electrons impact the detector at higher positions in z. All electrons are generated from a single point just above the sample surface in a conical distribution (half angle 2°).

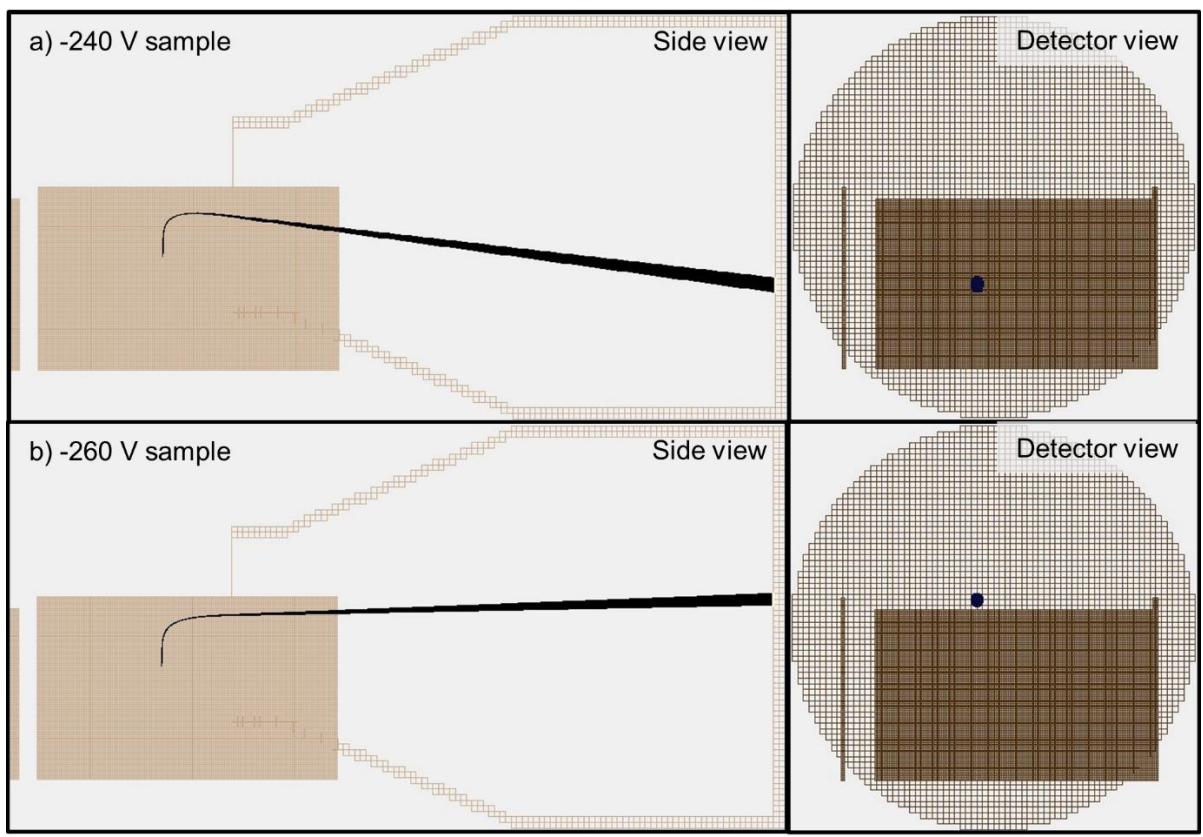

**Figure S3| Effects of sample voltage on SE distributions at the detector.** Side and detector views of 10 eV electrons traveling through our detector at (a) -240 and (b) -260 V on the sample electrode. The more negative potential effectively imparts more kinetic energy on the electrons that leave the shroud electrode resulting in the 10 eV electrons impacting the detector at a higher position. With -260 V on the sample electrode, 10 eV electrons impact the detector at the same position as 30 eV electrons with -240 V on the sample electrode.

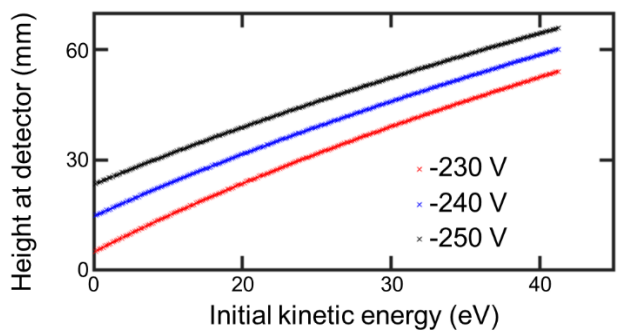

**Figure S4| Height at detector versus initial kinetic energy from SIMION simulations.** Simulations of electrons with 0 to 40 eV in kinetic energy were run with sample voltages from -230 V to -250 V. Electrons were generated just above the sample electrode and have velocities in the +Z direction. These distributions are perfectly described by the equation $H(E) = 122.5(E + E_0)^{0.5} - 65.75$ where $H(E)$ is impact height in mm (at scale) of an electron with initial kinetic energy $E$ and $E_0$ is a fitting parameter related to sample electrode voltage, which was fit to be ~33, 43 and 53 eV for sample electrode voltages of -230, -240, and -250 V, respectively. The -10 V changes between voltage and the corresponding 10 eV changes in $E_0$ suggest there is a -1 V: 1 eV correspondence between changes in sample electrode voltage and the extra energy imparted by the sample electron regarding impact height at the detector. Additionally, this suggests the sample electrode can be used to predictably control impact heights for energy spectroscopy.

Section S4: Effects of slit geometry on plume

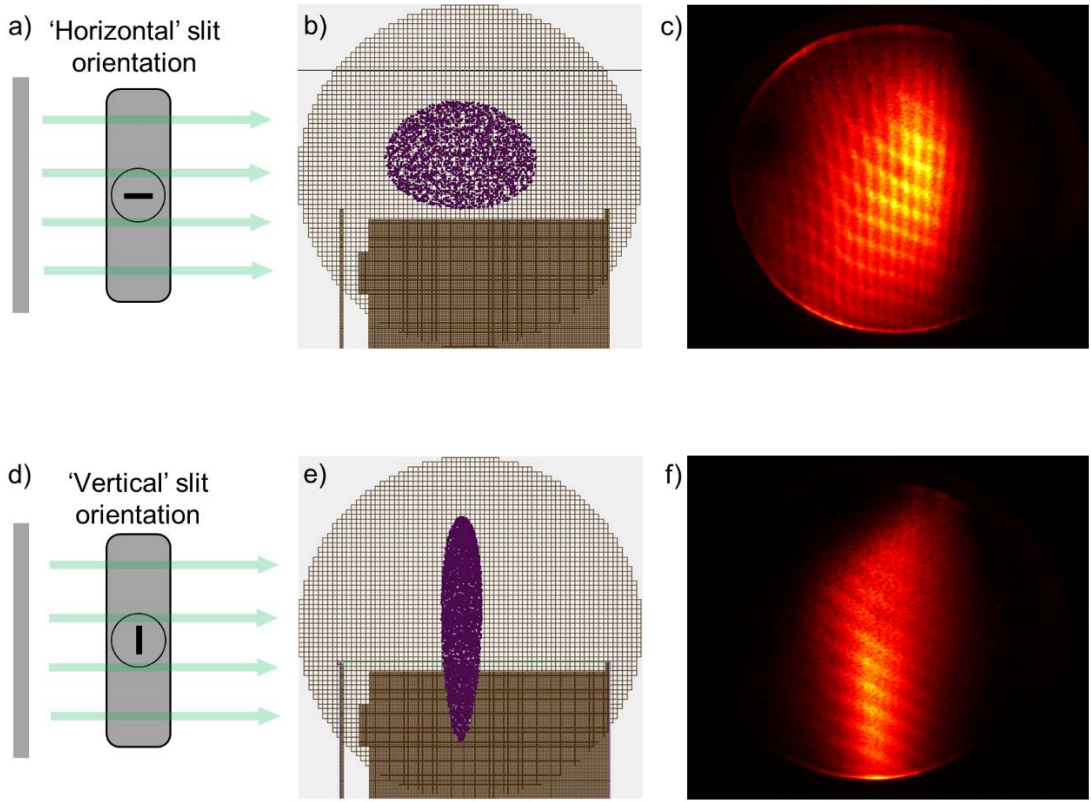

**Figure S5| Comparison of electron distributions for (a-c) 'horizontal' slit orientation and (d-f) 'vertical' slit orientation.** SIMION simulations and experimental results demonstrate that slit geometry on the shroud can impact plume shape on the detector with a 'horizontal' plume orientation compressing the SE plume along the energy axis and a 'vertical' plume compressing the SE plume along the angular/horizontal axis .(a) Schematic of a 'horizontal' slit orientation; this orientation was found to compress electron distributions in the energy dimension and consequently expand in the angular/horizontal dimension in both (b) SIMION simulations and (c) experimentally. (d) Schematic of a 'vertical' slit orientation; this orientation was found to expand electron distributions in the energy dimension and compress in the angular/horizontal dimension both in (e) SIMION simulations and (f) experimentally. SIMION simulations (b and e) were of 10 eV electrons generated just above the sample surface with velocities in a 'filled in' conical distribution (half angle 10°) centered along the +z axis. For these simulations, the slit geometry was at a 3 × 0.8 mm scale. Experimental data (c and f) are photographs of the SE plume on the detector screen, showcasing the experimentally observed shape change of the projected SE plume by slit geometry. (c) was taken with a sample electrode voltage of -165 V and (f) was taken with a sample electrode voltage of -170 V. The shape change is not due to clipping on the electrode, as verified in simulations, but instead likely is a result of lensing from the short axis of the slit (0.2 mm wide in these experimental data sets) which expands the SE plume along one dimension.

Section S5: Experiments on p-doped squares

Experiments were performed on Si wafers with varying patterns of p-doped (doped with B, ~$10^{19}$ cm$^{-3}$) and n-doped regions (doped with As, ~$10^{17}$ cm$^{-3}$), prepared according to a previous report.[1] Studies of the p-n interfacial regions (Fig. 2-4, S6-S9) were performed on regions of this sample containing p-doped squares surrounded by n-doped material. SEM micrographs were taken with the electron beam scanning with either 33.2 or 66.4 ms line times. Images of the plume across a line profile across the p-n interface (Fig. 3, S7, S8) were taken at 10 points along a ~40 μm line in spot mode on the SEM. This sample was also used for recording the SE spectra in Fig. 5e,f of n- and p- Si. Spectra were recorded while scanning at 25,000 × magnification in regions far away from p-n interfaces and other features that may impact the SE spectra. A separate, unpatterned p-type Si wafer was used for producing the three-dimensional plume plot and corresponding spectra in Fig. 5b-d. Images of the SE plume from this sample were recorded while scanning at 10,000 × magnification. Experimental parameters were optimized for each data set; electron optics voltages were set such that the SE plume was centered on the detector with a repeller voltage of ~ -3500 to -4000 V and a sample voltage around -180 to -200 V. In these cases, side electrodes were biased at ~± several hundred volts (one electrode biased, the other kept grounded). MCP/phosphor voltages were optimized for each experiment, i.e. higher gain (voltage) for data sets involving taking pictures of the detector or lower gain to prevent saturation of micrographs recorded with the plume and a PMT. Slit geometries—width and orientation—were optimized as well. Experiments on the lateral deflection used a 'vertical' slit to compress the SE plume along the x-dimension of the phosphor screen, and spectra were recorded with a 'horizontal' slit to compress the SE plume along the z-dimension. Micrographs were recorded with a 0.5 by 3 mm rectangular slit for increased flexibility in sample alignment due to the large slit sizes. Spectra were recorded using a smaller, and thereby more precise, 0.2 by 2 mm rectangular slit.

Micrographs were recorded of both a pristine, as received p-doped square, as well as a p-dope square that was coated in ~50 nm sputtered SiO$_2$ for identifying electric field effects in these structures even through a passivating layer (Fig. 4). Sputter coating was done at 180 W RF power, 5 mTorr reactant pressure, and 50 sccm flow rate, for 500 seconds. The p-doped squares, both uncoated and SiO$_2$ coated, were also analyzed by SEM imaging with an Everhart-Thorley detector (ETD), electron beam induced current (EBIC) and kelvin probe force microscopy (KPFM) (Fig. 4c-h). ETD images provided in Fig. 4c were recorded for the uncoated and coated samples concurrently with 33.2 ms line and with equivalent brightness and contrast settings time for direct comparison between the two samples. EBIC (Fig. 4e,f) was recorded of both samples using a Kleindick micromanipulator to position a probe (50 nm tip diameter W probe coated with Au) on the p-doped squares in both samples. The probe collected EBIC from the sample, which was amplified by a Keithley 428 transimpedance amplifier. In these measurements the SEM stage was grounded to the amplifier. Amplifier output was digitized to produce an EBIC image using a National Instruments USB-6363 multifunction DAQ and custom LabView code. Gain on the amplifier was $10^7$ V/A for the uncoated sample and $10^9$ V/A for the coated sample. To correct for differences in gain and allow direct comparison between images in Fig. 4g,f, all pixel values in the coated sample were divided by 100. There was essentially no EBIC when placing the probe on the oxide surface, as the insulating oxide layer prevented charge transfer from the sample to the probe.

Pressing the tip through the oxide layer restored EBIC signal (Fig. S6a) but doing so required breaking the oxide layer. Frequency-modulated KPFM (FM-KPFM) (Fig. 4g,h) was performed in a Bruker Dimension Icon using the PF-KPFM experiment module. An NSC18 Pt probe (nominal spring constant = 2.8 N/m) was used with a lift height of 30 nm for all measurements. Atomic force microscopy (AFM) data was concurrently performed, giving topological information on the analyzed area (Fig. S6b-c).

Line profiles (Fig. 4b,d,f,h) were taken horizontally across each image as indicated by red (uncoated sample) or blue ($SiO_2$ sputter coated sample) dashed lines over the micrographs (Fig. 4a,c,e,g). Colors, red and blue, of the line profiles correspond to the uncoated and coated samples, respectively. Line profiles were produced using custom MATLAB code that integrates horizontally across the image in a 10-pixel window and the line profiles in (i) were recorded using Gwyddion. A feature in the line profile in KPFM imaging of the coated sample (i, blue line profile) is a topological artifact resulting from the change in sample height from the p- to n-regions (Fig. S6b-c).

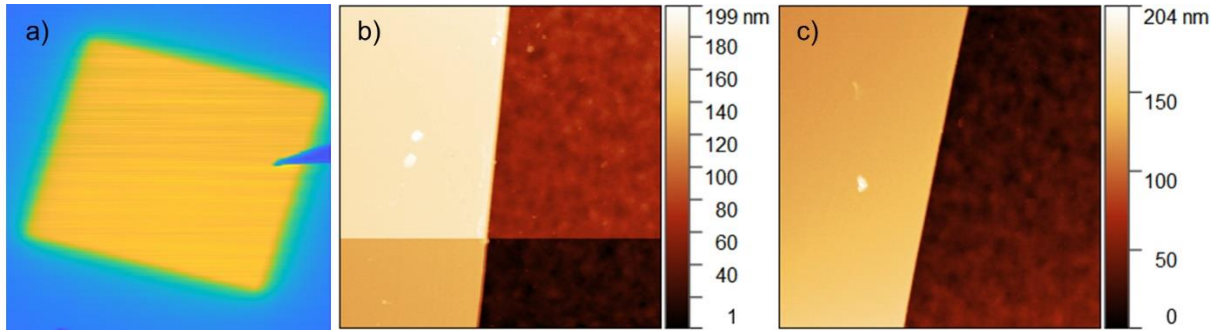

**Figure S6| Additional information from EBIC and KPFM experiments.** (a) EBIC image of the $SiO_2$ coated p-doped square after pushing the tip through the oxide layer. (b-c) AFM images of the sample areas analyzed with KPFM for the (b) uncoated and (c) coated p-doped squares, providing topographical information on the sample. There is a ~200 nm step down from the n-doped side (left of each image) to the p-doped side (right on each image).

Section S6: Imaging deflection of plume across p-n interfaces

Images of the SE plume were taken as a at ten points along a 40 μm line profile crossing p-n interfaces with opposite orientations, i.e. n- on the left/p- on the right and p- on the left/n- on the right, to observe the deflection caused by interfacial electric fields. The SEM was used in spot mode for this experiment. Images of the plume at each point in these line profiles and the ETD image of the interfaces are provided in Fig. S7. The horizontal centroid of the SE plume in each image was calculated using custom MATLAB code to produce Fig. 2g. Briefly, the code integrates the entire image along the vertical (energy) dimension to produce a vector of integrated counts versus horizontal position in pixels. The portion of this detector corresponding to the detector screen, 500 to 1300 pixels on the camera, was fit to a mixed Gaussian/Lorentzian peak to extract the center position, in pixels, of the SE plume. Distances along the line profile for each point were measured from point 1 using the ruler tool on the SEM control software, with the interface at ~ 20 μm from point 1. Center positions from the fit were plotted against distance from the interface along the line profile in Fig. 2g.

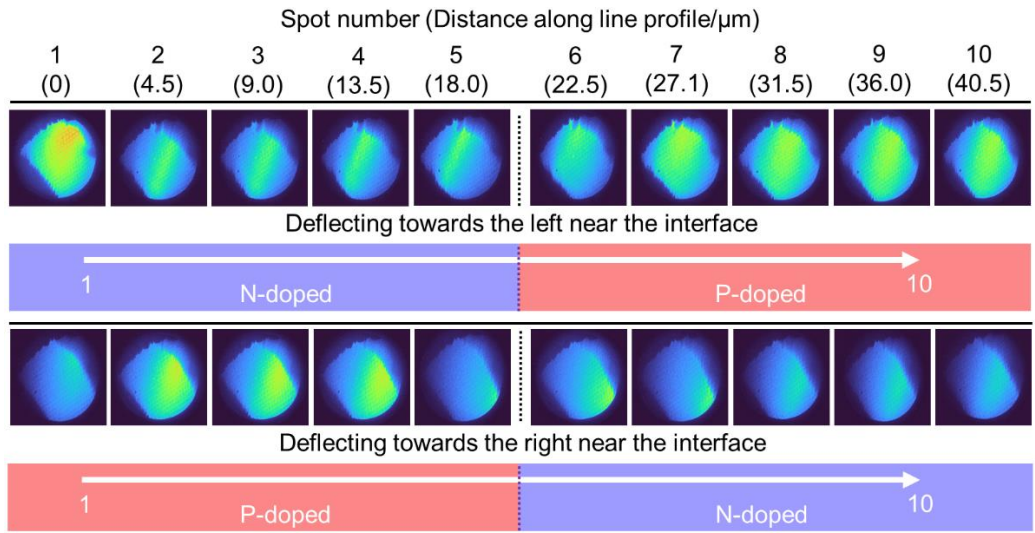

**Figure S7| Images of SE plumes on the detector taken across p-n interfaces.** Photographs were taken of the viewing screen at ten points along a line profile across a p – n junction. Sets of images were recorded with two p – n junction geometries: n-doped region on the left and p-doped region on the right (top) and the inverted geometry, i.e. p-doped on the left and n-doped on the right (bottom). These geometries are shown schematically in the bottom part of each data set. The center point of these distributions in the horizontal dimension, determined by peak fitting, was plotted versus distance to produce Fig. 2g.

A similar process was undertaken to investigate the energy dependence of the plume deflection due to interfacial electric fields (Fig. S8). Instead of integrating the full image in the vertical dimension, the portion of the image corresponding to the viewing screen (~400 – 850 pixels) was divided into four bins along the vertical dimension, from 426 to 825 with 100-pixel bin widths. This range was slightly truncated from the full screen diameter to remove edge effects at the extremes. These four integrated vectors were analyzed in the same way as in Fig. 2g to extract center position of the SE plume in the horizontal dimension versus position along line profile. In this analysis, the bottom two quartiles correspond to lower energy electrons and the top two quartiles correspond to higher energy electrons. Plotting deflection versus position on the line profile for these four quartiles (Fig. S8) shows that deflections are larger for lower energy electrons than higher energy electrons, suggesting lower energy electrons are more perturbed by lateral electric fields.

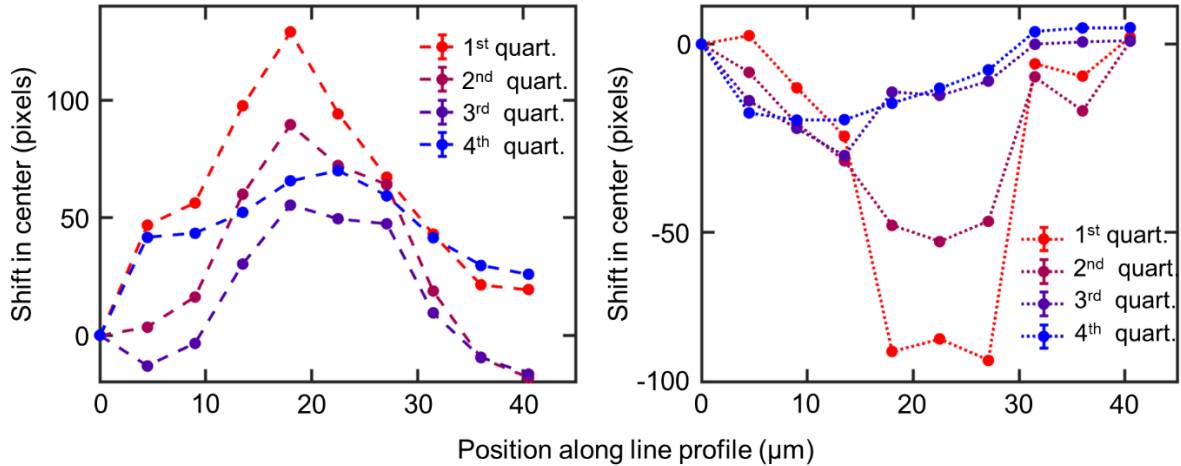

**Figure S8| SE plume deflection for different energy electrons.** The plume images (Fig. S7) were divided into quartiles along the energy dimension and the horizontal center of the SE plume were determined for each quartile at all 10 points along the line profile. The 1$^{st}$ quartile corresponds to the lowest energy electrons and the 4$^{th}$ quartile corresponds to the highest energy electrons. Left plots are from crossing a p-n junction with n-doped material on the left, and the right plots are from crossing a p-n junction with p-doped material on the left. Lower energy electrons experience larger deflections than higher energy electrons, as evidenced by the generally larger shifts in plume center for the 1$^{st}$/2$^{nd}$ quartile than the 3$^{rd}$/4$^{th}$ quartile. In each image, the interface is at ~20 μm along the line profile. Shifts were defined as differences in center position for the fits of each point relative to the center at position 1 (0 μm). Error bars representing propagated errors from each fit are present but are too small to be seen on these images. quart.; quartile.

Section S7: Spatially filtered phosphor screen for imaging sides of p-doped square

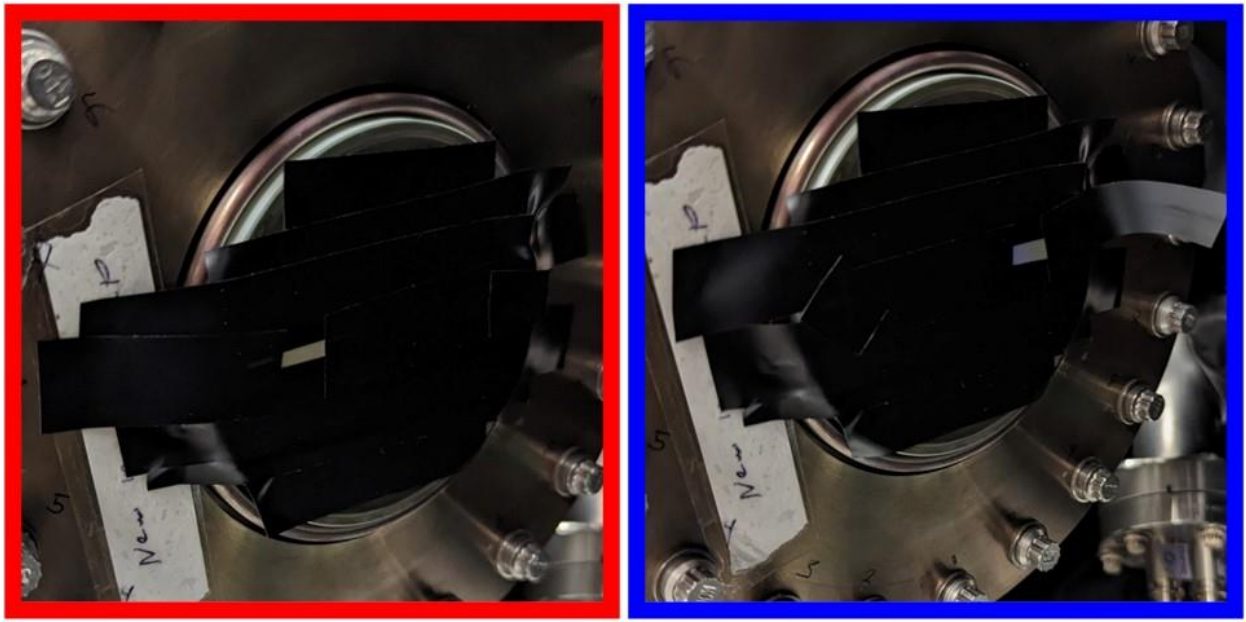

**Figure S9| Spatially filtered phosphor screen.** Used to record micrographs of the left (red) and right (blue) sides of a p-doped square in Fig. 3d,e. Colors in these images correspond to panels in Fig. 3.

Section S8: Selectively imaging top and bottom sides of p-doped squares

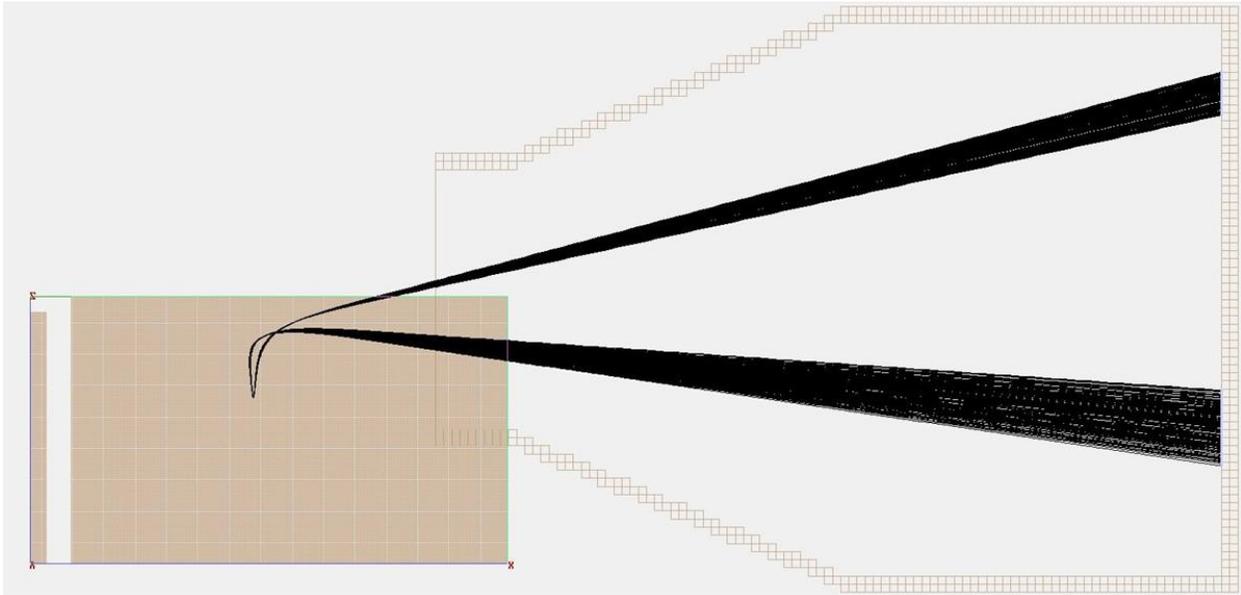

**Figure S10| Schematic for imaging top and bottom sides of p-doped square.** SIMION Simulation of electron trajectories through our detector with initial momenta going towards or away from the detector screen. Electrons initially moving towards the detector screen (right in this view) impact the detector at higher positions in $z$, and electrons initially moving away from the detector screen (left in this view) impact the detector at lower positions in $z$. Leveraging this, we were able to selectively image the top and bottom of a p-doped square with a central slit (Fig. 3f,g). By making the sample electrode voltage more negative, the SE plume moves up in $z$, allowing selective imaging of electrons deflected away from the detector screen due to interfacial electric fields at the top side of the p-doped square (Fig. 3f). Conversely, by making the sample electrode voltage more positive the SE plume moves down in $z$, allowing selective imaging of electrons deflected towards the due to interfacial electric fields at the bottom side of the p-doped square (Fig. 3g). Simulations were performed with -240 V sample electrode voltage and -3000 V repeller electrode voltage. Electrons were generated with conical distributions with 10° half angle centered along $<x, y, z> = <0.5, 0, 1>$ or $<-0.5, 0, 1>$, for electrons starting towards and away from the detector screen, respectively.

Section S9: Secondary Electron Spectroscopy

Full SE plume plots (Fig. 5) were collected by taking a series of images of the plume on the detector using a digital camera while scanning the sample electrode voltage. A full SE plume for the p-type Si with a 'horizontal' slit orientation (Fig. 5a-d) was produced by scanning from -205 to -125 V, scanned in 1 V steps. This set of images was converted into 3D plume plots using custom MATLAB code. Briefly, the code integrates a portion of the center of the axis of SE plume on each image; a rectangular window 600 pixels in the horizontal dimension and 50 pixels in the vertical dimension was selected and integrated along the vertical dimension to produce a vector of horizontal position, in pixels, versus integrated counts. This process is shown schematically in Fig. S11. Integration was repeated for all images in the full set, using the same position on the camera for all images. All vectors were combined to produce the 3D plume plots (Fig. 5b) by plotting integrated counts ($z$-axis in 3D plot) versus sample voltage ($x$-axis in plot) and horizontal position ($y$-axis). Energy (Fig. 5c) and horizontal/momentum (Fig. 5d) spectra were produced using the 3D plume plot (Fig. 5b) by integrating the red and blue rectangles in the inset in Fig. 5b along the horizontal and sample voltage axes, respectively.

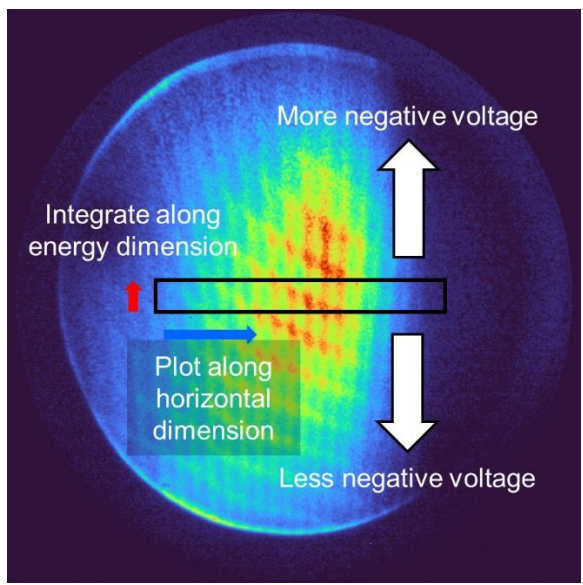

**Figure S11| Schematic of analysis to produce 3D plume plots.** Changing sample electrode voltage can be used to control position of the SE plume on the detector in the vertical dimension (Fig. 5a, S3), with more negative/positive voltages moving the SE plume up/down. This was exploited to record a series of images of the SE plume on the viewing screen as the sample electrode voltage was scanned. In each image, a rectangular window was integrated along the energy/vertical axis (red arrow) to produce a vector of integrated counts versus horizontal position (blue arrow) for each sample electrode voltage. The 3D plumes were produced from this set of vectors, plotting integrated counts versus horizontal position and sample electrode voltage. The mesh pattern observed is a projection of the Cu mesh on the mu metal ToF tube, which we can use for directly measuring the size of the SE plume as it enters the mu metal ToF tube or can act as a known grid pattern that we can use for deconvolving artifacts from the detector response from the projected image of the SE plume on the viewing screen.

Spectra of n- and p-doped Si (Fig. 5e,f) were acquired in the same manner as the spectrum in Fig. 5c, but instead using a 20-pixel integration window in the vertical dimension on the images of the SE plume to produce the initial 3D plume plot. The 3D SE plumes were integrated along the horizontal dimension to produce spectra of integrated counts versus sample electrode voltage. These spectra were fit to a mixed Gaussian/Lorentzian peak shape to extract peak fitting parameters (height, center, etc.) (Table S2).

**Table S2| Fitting parameters from mixed Gaussian/Lorentzian peak fitting to SE spectra in Fig. 5e,f.** Fit value ± standard deviation from the fit.

| Material | p-type | n-type |
| --- | --- | --- |
| Peak position (V) | $-196.60 \pm 0.04$ | $-194.70 \pm 0.03$ |
| Peak height | $1590000 \pm 2000$ | $934300 \pm 1200$ |
| Peak width (V) | $33.0 \pm 0.2$ | $36.49 \pm 0.09$ |
| Peak Area | $55600000 \pm 200000$ | $34470000 \pm 110000$ |

Of note, spectra recorded using our electron detector do not quantitatively agree with published results. Namely, recorded spectra were very symmetric whereas prior results indicate asymmetric spectra with tails at high SE energies. One contributor to this discrepancy is intrinsic connection between angular spread and spread over the z-dimension of electrons of the SEs projected onto the detector (Fig. 1c). Another variable that affects the shape of the projected SE energy spectrum is the size and shape of the slit in the shroud electrode (Fig. S5). Simulations of detector response using a published SE energy spectrum for p-type Si[2] indicate that the detector response results in peak broadening and shape change (Fig. S12-S14). With no angular spread (Fig. S12b,c), the $z$-position of electrons impacting the detector agree in shape with the input spectrum (Fig. S12a) and in full-width-tenth-max (FWTM), determined by calibrating $z$-position versus electron energy, 1.2 eV per graphic unit based on the shift in average $z$-position of electrons impacting the detector from a 10 V change in sample electrode voltage from -240 to -250 V. In these simulations, 1 graphics unit is equivalent to 1 mm at scale. The FWTM in energy was then calculated by estimating the FWTM in graphics units and multiplying by this calibration constant Adding an angular component by assigning the electrons to be generated with initial velocities in 'filled in' conical distributions, causes a change in peak shape and peak broadening. For a 10° (half angle) conical spread (Fig. S12d,e), the FWTM of the electrons on the detector is ~20 eV, and for a 30° (half angle) conical spread (Fig. S12f,g), the FWTM of the electrons on the detector is ~50 eV. This is similar to the FWTM recorded experimentally for p-type Si, ~53 V (Fig. 5c). With a 40° cone, the energy spectrum is well replicated by a histogram of electron positions along the $z$-axis (Fig. S13). Thus, the angular distribution of the SEs generated from this material is well represented as a conical distribution with a half angle of ~30 – 40°. Electrons generated with initial trajectories close in angle to the +z axis contribute to the center position of the impact heigh distribution and those generated with initial trajectories far from the +z axis contribute to the extreme ends of the impact height distribution (Fig. S14). SEs may be generated at higher angles as well but contribute less to the recorded spectrum. Furthermore, it is possible that the SE angular distribution is also a function of SE energy, i.e. different SE energies may require different size cones to fully describe the SE energy spectrum. These results demonstrate that the convolution between angular and energy spreads of the SEs result in peak broadening and shape changes between prior published SE spectra and the energy spectrum recorded by our SE detector, and that deconvolutions can be used to extract energy and angular information from the SE plume.

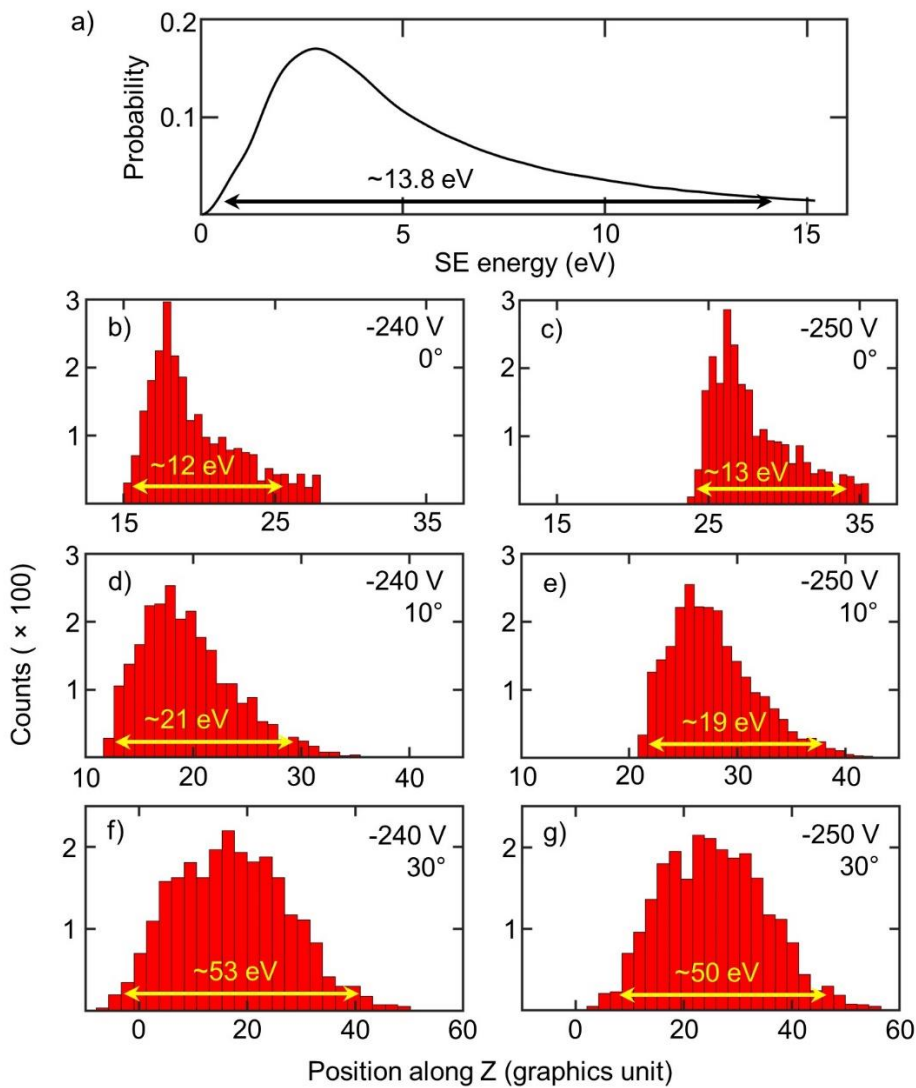

**Figure S12| Distributions of electron heights at detector using different angular spreads and published SE energy spectrum.** The distribution of electron heights (in Z/along the energy axis) was investigated at varying angular distributions using SIMION. Electrons were generated just above the sample surface in our detector with initial velocities in a "filled in" conical distribution with half angles ranging from 0° – 30° centered along the +$z$ direction and with energies randomly assigned from a probability distribution (a) produced from published SE spectrum of p-doped Si.[2] The $z$-positions of the electrons impacting the detector are plotted in histograms in (b-g); with a sample electrode voltage of -240 V and cone half angles of (b) 0°, (d) 10°, and (f) 30°, and with a sample electrode voltage of -250 V and cone half angles of (c) 0°, (e) 10°, and (g) 30°. In total, each histogram contains 2500 electrons. The full width tenth max (FWTM) energy spreads of all plots are provided. For the histograms, FWTM in energies were estimated by estimating FWTM in $z$-position distributions and multiplying by a calibration constant of 1.2 eV per graphics unit, determined as the 10 eV change in electron energy, due to a 10 V change in sample electrode voltage, divided by the change in average $z$-position between electrons at the two different sample voltages. Increasing the cone half angle results in a drastic change in spectrum shape and broadness. At 0° (b and c) the $z$-distribution histogram displays roughly the same shape as the input spectrum, i.e. asymmetric with a tail at higher $z$-positions. FWTM is similar for these distributions to the input spectrum. Conversely, a 30° cone (d and g) is much broader in FWTM and is more symmetric in peak shape. This broadening and shape change likely results in the observed SE spectrum recorded with our detector (Fig. 5c). Simulations were performed with a -3000 V repeller electrode voltage.

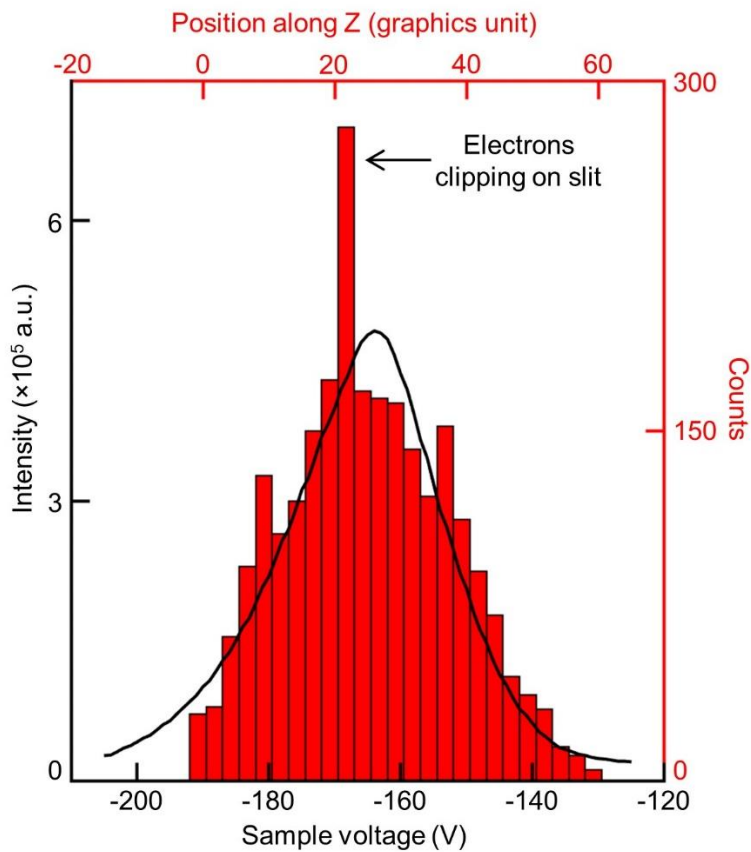

**Figure S13| Distributions of electron heights at detector using a 40° angular spread (red) with an overlain experimental spectrum (black).** The distribution of electron heights (in *z*/along the energy axis) was investigated using SIMION. Electrons were generated just above the sample surface in our detector with initial velocities in a "filled in" conical distribution with a half angle of 40° centered along the +*z* direction and with energies randomly assigned from a probability distribution in Fig. S12a. The histogram of electron heights (positions along Z) well replicates the experimental SE energy spectrum of p-type Si (Fig. 5c), further suggesting electrons are generated at 30 – 40° angles from the normal of the surface. Of note, at this angle there is substantial clipping of electrons on the slit of the shroud electrode (*z* ~ 20 graphics units), which was observed in experiments, particularly when the sample location of interest was off centered in the slit. Further, some electrons impacted the bottom of the mu metal ToF tube in this simulation, resulting in the sharp drop in counts below ~0. This was not an issue in the experimental spectrum since it was recorded by moving the SE plume along the z-axis with the sample electrode voltage. Simulations were performed with a -3000 V Repeller electrode voltage and a -250 V sample electrode voltage. This figure is replicated from Fig. 5c.

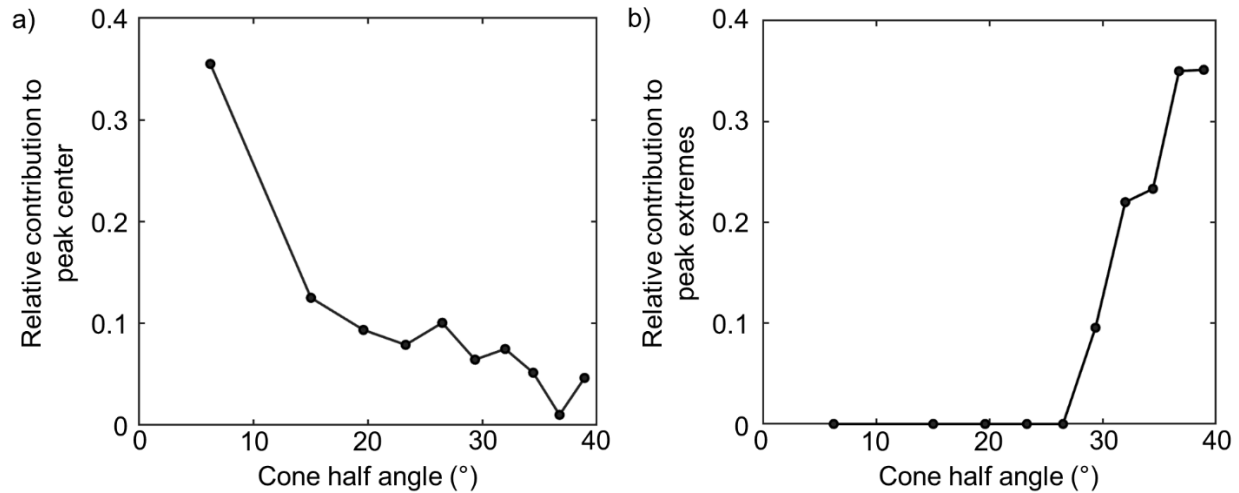

**Figure S14| Relative contribution to peak center in Fig. S13 from electrons generated at different angles from the +z axis.** Impact heights for electrons simulated to produced Fig. S13 were analyzed by initial angle of trajectory relative to perfectly vertical (parallel to the +z axis). Specifically, these electrons were binned into 10 separate concentric cones across the full 40° conical distribution, such that electrons coming out with nearly vertical or far from vertical could be compared based on small or large angles relative to +z, respectively. Bin sizes were set such that each contained roughly the same number of electrons. These impact heights were further binned into 20 equal size bins across the full range of impact trajectories in Fig. S13. (a) The peak center in the binned distribution is around ~27 graphics units. The relative contributions to this peak center from all angular bins were calculated by summing the electron counts for each angular bin across the 3 center bins in the impact height distribution, which roughly encompass 23 to 32 graphics units on the impact height distribution and dividing by the total number of electrons impacting in these bins across all angles. These relative contributions are plotted verses cone half angle in this plot. Unsurprisingly, electrons generated at small angles (relative to the +z axis) contribute more to the center of the impact height distribution than electrons generated at large angles. (b) Likewise, repeating this process for electrons impacting at the extreme ends of the impact height distribution (~ -2 or +61 graphics units) reveal these electrons are mostly from those generated at high initial angles, as high angles result in substantial broadening along the energy axis. Bins used for producing the plot in (b) were of impact heights of ~-2 to 3 and ~53 to 61 graphics units. Only electrons generated at angles greater than ~27° contribute to the extreme impact heights.

References:

(1) Anderson, M. L.; Nakakura, C. Y.; Kellogg, G. L. *Imaging doped silicon test structures using low energy electron microscopy*; SAND2009-7981; OSTI.gov, 2010. https://www.osti.gov/servlets/purl/993607.
(2) Srinivasan, A.; Han, W.; Zheng, M.; Khursheed, A. Characterization of materials using the secondary electron energy spectromicroscopy technique. *Optical Materials: X* **2021**, *12*, 100121. DOI: https://doi.org/10.1016/j.omx.2021.100121.